\documentclass[12pt]{article}
\usepackage[colorlinks,citecolor=blue,urlcolor=blue,filecolor=blue]{hyperref}
\usepackage{apacite}
\usepackage[longnamesfirst,sort]{natbib}

\title{A Nonparametric Bayesian Item Response Modeling Approach for Clustering Items and Individuals Simultaneously}

\author
{Guanyu Hu \thanks{Department of Statistics, University of Connecticut; Department of Statistics, University of Missouri Columbia}~~ Zhihua Ma \thanks{Corresponding Author, Email: mazh1993@outlook.com, College of Economics, Shenzhen University}~~ Insu Paek\thanks{Department of Educational Psychology \& Learning Systems, Florida State University}}
\date{}

\usepackage{rotating, hyperref}
\usepackage{mathrsfs}
\usepackage{url}
\usepackage[margin=1in]{geometry} 
\usepackage{graphicx,subfigure}
\usepackage{mathtools}

\usepackage{setspace}
\usepackage{amsmath,amsthm,amssymb,bm}
\usepackage{float}
\usepackage{cases}
\usepackage{algorithm}
\usepackage{algorithmic}
\usepackage{multirow}
\usepackage{color}
\usepackage{siunitx}
\usepackage{xcolor}
\usepackage{amsfonts}
\usepackage{booktabs}
\usepackage{siunitx}
\usepackage{pdfpages}

\begin{document}
\maketitle
\vspace*{-1cm}
\begin{abstract}
Item response theory (IRT) is a popular modeling paradigm for measuring subject latent traits and item properties according to discrete responses in tests or questionnaires. There are very limited discussions on heterogeneity pattern detection for both items and individuals. In this paper, we introduce a nonparametric Bayesian approach for clustering items and individuals simultaneously under the Rasch model. Specifically, our proposed method is based on the mixture of finite mixtures (MFM) model. MFM obtains the number of clusters and the clustering configurations for both items and individuals simultaneously. The performance of parameters estimation and parameters clustering under the MFM Rasch model is evaluated by simulation studies, and a real date set is applied to illustrate the MFM Rasch modeling.
\end{abstract}

\section{Introduction}\label{sec:intro}

In psychometrics, item response theory (IRT) is a popular modeling paradigm for measuring subject latent traits and item properties according to discrete responses in tests or questionnaires. In IRT models, the discrete responses are explained as a function of respondents' latent traits and item properties. The Rasch model \citep[][]{rasch1960}, one of the most widely used IRT models, can be used to infer the latent abilities of the test-takers as well as the item difficulties. Generally, an assumption of most IRT models including the Rasch model is measurement invariance, which means all items measure the latent trait in the same way for all subjects, that is, the item response function for an item is invariant across all subjects. However, this assumption may be violated in some cases by poorly-constructed items or items with varying difficulty levels for different subgroups of persons. 

Heterogeneity in IRT analysis is typically addressed by latent class analysis (LCA) approaches and mixture IRT models. 
The LCA model assumes that the sample of respondents is drawn from a population divided into several latent classes, with individuals in the same class sharing the same distribution of the response variables \citep[][]{lazarsfeld1968latent}. LCA model attempts to identify distinct classes of individuals and classify the test-takers. For an overview of LCA methods and applications, see e.g. \cite{hagenaars2002applied}.
The mixture IRT models are regarded as a combination of IRT models and LCA approaches, whose basic idea is that the same IRT model holds within subjects in the same class, whereas different IRT models hold among different classes \citep[][]{gnaldi2016multilevel}. Thus, with mixture IRT models, it is possible to identify distinct subpopulations within a larger population, each of which responds differently to a set of items \citep[][]{rost1990rasch, rost1997applications}.

Under both the LCA and mixture IRT frameworks, however, the number of clusters should be pre-specified by the user before obtaining the estimates.
In some applications, the number of clusters is chosen by prior knowledge, which is not always available. Another approach for selecting the number of clusters is based on information criteria such as the
Akaike information criterion \citep[AIC;][]{akaike1998information, bartolucci2007class}, and Bayesian information criterion \citep[BIC;][]{schwarz1978estimating}, which are computationally ineffective and time consuming \citep[][]{pan2014bayesian}. Moreover, the two-stage approach that inferences conditional on the specification of the number of clusters ignores the uncertainty in the selection process \citep[][]{yang2011nonparametric}.

This problem is circumvented by nonparametric Bayesian approaches. Nonparametric Bayesian approaches allow joint inferences on the clustering information, which includes both the number of clusters, clustering configurations, and parameters on individuals and items simultaneously. For example, \citet{miyazaki2009bayesian} proposed a Bayesian nonparametric mixture IRT model using a Dirichlet process prior. \citet{liu2019three} developed a Bayesian nonparametric ordered latent class model with a modified Chinese restaurant process prior. Although \cite{bartolucci2017nonparametric} used a Bayesian nonparametric method on IRT model with a LCA formulation, the number of clusters was still determined via model selection criteria. 
Research on Bayesian nonparametric approaches for dealing with clustering in IRT models are still limited, and the existing approaches mainly focus on Dirichlet process prior.

The aim of this paper is to propose a new hierarchical Rasch model for simultaneously clustering items and individuals under a Bayesian framework. Bayesian nonparametric approaches offer choices to allow uncertainty in the number of clusters, and provides an integrated probabilistic framework under which the number of clusters, the clustering configuration, subject latent abilities and item difficulties are simultaneously estimated.
For the clustering problem, Bayesian
inference provides a probabilistic framework for simultaneous inference of the
number of clusters and the clustering configurations based on a Dirichlet process
mixture model \citep[DPM;][]{ferguson1973bayesian}. \cite{miller2013simple}
points out that DPM will produce extremely small clusters and make the
estimation for number of clusters inconsistent. The mixture of finite mixtures (MFM)
model proposed by \cite{miller2018mixture} provides a remedy to over-clustering problem for Bayesian nonparametric methods. MFM is widely applied in different areas such as regional economics \citep{hu2020bayesian}, social science \citep{geng2019probabilistic} and environmental science \citep{geng2019bayesian}.
Thus, our key idea is to formulate the basic Rasch model into a hierarchical form, and assign the MFM priors for the latent ability and item difficulty parameters. 
By assigning the MFM priors, our proposed approach allows for the uncertainty in the number of clusters, and through a Bayesian framework,  estimation is obtained for both the number of clusters, clustering configurations and parameters in terms of person latent traits and item difficulty parameters. Hence, the proposed method can be effectively used to identify item and individual clustering. Test developers may use the clustering results of item difficulties for their test construction. In addition, the information on the clustering of subject latent abilities could be useful when grouping of individuals is needed for placement or other diagnostic assessment purposes.

The remainder of the paper is organized as follows. In
Section~\ref{sec:method},
we present the proposed hierarchical Rasch model with MFM. In
Section~\ref{sec:bayesian_inference}, the Bayesian inference procedure including prior and posterior distributions and model comparison criterion are discussed. Simulation studies and the results are described 
in
Section~\ref{sec:simu}. For the purpose of illustration, our proposed methodology is applied
to
an English test data in Section~\ref{sec:real_data}. Finally, we conclude this
paper with a discussion in Section~\ref{sec:discussion}.

\section{Methodology}\label{sec:method}
\subsection{Item Response Model}
Suppose there are a total of $N$ subjects and $J$ items, and $\bm{y}_i = (y_{i1}, \cdots, y_{iJ})'$ represents the binary response vector for the $i$th individual at $J$ items. $y_{ij} = 1$ means correct answer at the $j$th item for individual $i$, while $y_{ij}=0$ represents wrong answer. Denoting $P(y_{ij} = 1|\theta_i) = p_{ij}$, the Rasch model \citep{rasch1960} is given as
\begin{equation} \label{eq: Rasch}
     p_{ij} = \frac{\text{exp}(\theta_i - b_j)}{1 + \text{exp}(\theta_i - b_j)},
\end{equation}
for $i = 1, \cdots, N$, $j = 1, \cdots, J$, where $\bm{\theta} = (\theta_1, \cdots, \theta_N)$ is the vector of latent abilities, and $\bm{b} = (b_1, \cdots, b_J)$ denotes the vector of item difficulty parameters. 

Under a Bayesian framework, the Rasch model in \eqref{eq: Rasch} is written as 
\begin{equation}
\begin{split}
		y_{ij} &\sim \text{Bernoulli}(p_{ij}),\\
	\text{log}(p_{ij}/(1-p_{ij})) &= \text{logit}(p_{ij}) = \theta_i - b_j.	
\end{split}
\label{eq:rasch_model}
\end{equation}

By giving the prior distributions of the unknown parameters and through the Markov chain Mote Carlo (MCMC) sampling algorithm, the posterior estimates of parameters for $i = 1, \cdots, N$, $j = 1, \cdots, J$ can be obtained.

%

\subsection{Dirichlet Process}\label{sec:DP}
Next, we introduce nonparametric Bayesian methods for clustering the individuals and items.
Dirichlet process (DP) is an universally used Bayesian nonparametric method for capturing heterogeneity effects of the data. DP is applied
to Rasch model for capturing heterogeneity on both item difficulties and person abilities. In this part for simplicity, we discuss the grouping methods for abilities $\theta_1,\cdots,\theta_N$. Assume we have Rasch model as \eqref{eq: Rasch}, the prior distribution for $\theta_i,i=1,\cdots,N$ is the Dirichlet process with concentration parameter $\alpha$ and base distribution $G_0$ as below:
\begin{equation}
	\begin{split}
		&\theta_i\sim G\\
		&G\sim \text{DP}(\alpha,G_0).
	\end{split}
	\label{eq:dp}
\end{equation}
Since the DP yields distributions $G$ that are almost surely discrete \citep{ferguson1973bayesian}, the distinguished values among $\theta_1,\cdots,\theta_N$ induce a grouping among the abilities. The observations $i$ and $i'$ are in the same group when $\theta_i=\theta_{i'}$. For observation $i$, we have its grouping allocation label $z_i \in\{1,\cdots,K\}$, and the ability of observation $i$ is denoted by $\theta_{z_i}$, where $K$ is the total number of groups. In this way, we have an equivalent model as 
\begin{align}
	\begin{split}
		z_i|\bm{\pi}&\sim \text{Discrete} (\pi_1,\cdots,\pi_K),\\
		\theta_k &\sim G_0,\quad k=1,\cdots,K,\\
		\bm{\pi} &\sim \text{Dirichlet} (\alpha/K,\cdots,\alpha/K),
	\end{split}
	\label{eq:dpmm}
\end{align}
where $\bm{\pi}=(\pi_1, \cdots, \pi_K)$, $K\rightarrow \infty$ \citep{ishwaran2001gibbs}. 

The distribution of $z_i$ is marginally given by a stick-breaking construction \citep{sethuraman1994constructive} of DP as 
\begin{equation}
	\begin{split}
		z_i&\sim \sum_{h=1}^\infty \pi_h \delta_h,\\
		\pi_h&=\nu_h\prod_{\ell\leq h}(1-\nu_\ell),\\
		\nu_h&\sim \text{Beta}(1,\alpha),
	\end{split}
	\label{eq:s-b_construct}
\end{equation}
where $\delta_h$ is the Dirac function with mass at $h$. Based on the stick-breaking construction of DP, a MCMC algorithm \citep{ishwaran2001gibbs} can be used for parameter estimation. However, DP results in producing extraneous clusters in the posterior, leading to
inconsistent estimation on~$K$, the number of clusters, even when the
sample
size goes to infinity.

\subsection{Mixture of Finite Mixtures Model}\label{sec:MFM}
 \cite{miller2013simple} showed that the
posterior distribution on the number of clusters does not converge to the true
number of components under DP. \cite{miller2018mixture} proposed a modification of the
Chinese restaurant process (CRP) \citep{pitman1995exchangeable}, called a mixture of finite mixtures (MFM) model, to circumvent this issue: 
\begin{eqnarray}\label{eq:MFM}
\begin{split}
K & \sim p(\cdot), \\
(\pi_1, \ldots, \pi_K) \mid K &\sim \text{Dirichlet}(\gamma, \ldots, \gamma), \\
 z_i \mid K, \bm{\pi} & \sim \sum_{k=1}^K  \pi_k \delta_k, \quad  i=1, \ldots, N, 
\end{split}
\end{eqnarray}
where $p(\cdot)$ is a proper probability mass function (p.m.f) on $\{1, 2,
\ldots, +\infty\}$ and $\delta_k$ is a
point-mass at $k$. A default choice of $p(\cdot)$ is a $\mbox{Poisson}(1)$
distribution truncated to be positive \citep{miller2018mixture}, which is
assumed through the rest of the paper. Like the stick-breaking representation
in \eqref{eq:s-b_construct} of Dirichlet process, the MFM also has a
similar construction. If we choose $(K-1) \sim \mbox{Poisson}(\lambda)$ and
$\gamma=1$ in \eqref{eq:MFM}, the mixture weights~$\pi_1,\cdots,\pi_K$ can be
constructed as follows:
\begin{enumerate}
	\item Generate $\eta_1,\eta_2,\cdots \overset{\text{iid}}{\sim}
\text{Exp}(\lambda)$,
	\item $K=\min\{j:\sum_{k=1}^j \eta_k \geq 1\}$,
	\item $\pi_k=\eta_k$, for $k=1,\cdots,K-1$,
	\item $\pi_K=1-\sum_{k=1}^{K-1}\pi_k$.
\end{enumerate}
The Gibbs samplers are easily constructed in stick-breaking framework \citep{ishwaran2001gibbs} for MFM. For ease of exposition, we refer MFM as $\text{MFM}(\gamma,\lambda)$ with default choice of $p(\cdot)$ being $\mbox{Poisson}(\lambda)$.

\subsection{Hierarchical Rasch Model with MFM}\label{sec:hierachical}

In order to allow for simultaneously heterogeneity detection of person and item parameters, the MFM process in Section~\ref{sec:MFM} is introduced for the parameters $\bm{b}$ and $\bm{\theta}$ in the item response model. The hierarchical Rasch model with MFM is given as follows, for $i = 1, \cdots, N$, $j = 1, \cdots, J$:
\begin{equation}
	\begin{split}
			y_{ij} &\sim \text{Bernoulli}(p_{ij}),\\
	\text{log}(p_{ij}/(1-p_{ij})) &= \text{logit}(p_{ij}) = \theta_{z_i} - b_{g_j},\\
	b_{g_j} &\sim N(0, \psi_b^{-1}), \\
	\theta_{z_i} &\sim N(0, \psi_{\theta}^{-1}),\\
	\psi_{\theta} &\sim \text{Gamma}(100, 1),\\
	g_j|K_b, \pi_b, \lambda_b &\sim \text{MFM}(\gamma,\lambda_b),\\
	z_i|K_{\theta}, \pi_{\theta}, \lambda_{\theta} &\sim \text{MFM}(\gamma,\lambda_\theta),
	\end{split}
	\label{eq:hierachial_model}
\end{equation}
where $\psi_b^{-1}=0.01$, $\bm{g}=(g_1, \cdots, g_J)$ and $\bm{z}=(z_1, \cdots, z_N)$ denote the vectors of cluster configurations of parameters $\bm{b}$ and $\bm{\theta}$, respectively. 
The value of $\psi_{b}$ and the prior for $\psi_{\theta}$ are set for mean controlling purpose. 
MFM denotes the clustering method introduced in Section~\ref{sec:MFM}. 

\section{Bayesian  Inference}\label{sec:bayesian_inference} 
\subsection{Prior Specification and Posterior Distribution}
For the hierarchical Rasch model with MFM introduced in Section~\ref{sec:hierachical}, the set of parameters is denoted as $\Theta =\{(\theta_{z_i}, b_{g_j}, K_b, K_\theta, \pi_b, \pi_\theta,  \lambda_b, \lambda_{\theta}): i = 1, \cdots, N, j = 1, \cdots, J\}$. Priors for the hyperparameters are $\lambda_b \sim \text{Log-Normal}(0,1)$ and $\lambda_{\theta} \sim \text{Log-Normal}(0,1)$. With the prior distributions specified above, the posterior distribution of these parameters based on the data $D = \{(y_{ij}): i = 1,\cdots, N, j = 1, \cdots, J\}$ is given by
\begin{align*}
	\pi(\Theta|D) &\propto L(\Theta|D) \pi(\Theta)\\
	& \prod_{i=1}^{n} \prod_{j=1}^{J} f(y_{ij}|b_{g_j}, \theta_{z_i}) f(b_{g_j})f(\theta_{z_i})f(g_j)f(z_i) \pi(\Theta).
\end{align*}
The analytical form of the posterior distribution of $\Theta$ is unavailable. Therefore, we carry out the posterior inference using the MCMC sampling algorithm to sample from the posterior distribution and then obtain the posterior estimates of the unknown parameters. The estimated clustering allocations $g_j$'s and $z_i$'s are obtained from the posterior modes of the MCMC samples, while the other parameter estimations are obtained through posterior means of the MCMC samples. Computation is facilitated by the \textbf{nimble}\citep{de2017programming} package in \textsf{R} \citep{Rlanguage2013}, which uses syntax similar to \textsf{WinBUGS} \citep{lunn2000winbugs} and \textsf{JAGS}\citep{plummer2003jags}, but generates \textsf{C++} code for faster computation.

\subsection{Bayesian Model Comparison}
In the hierarchical Rasch model with MFM, the prior distributions of $\lambda_b$ and $\lambda_{\theta}$ have several choices, including the Gamma distribution, Uniform distribution and Log-normal distribution. In order to choose the most suitable priors for these parameters, model comparison criteria under the Bayesian framework are used. One of the most commonly used criterion is the Deviance Information Criteria \citep[DIC;][]{spiegelhalter2002bayesian}, which measures the fitness as well as the complexity of the model. The DIC is defined as 
\begin{gather}
	\text{DIC}=\text{Dev}(\bar{\Theta})+2p_D,
\end{gather}
where $\text{Dev}(\bar{\Theta})$ is the deviance function,
$p_D=\overline{\text{Dev}}({\Theta})-\text{Dev}(\bar{\Theta})$ is the effective
number of model parameters, $\bar{\Theta}$ is the posterior mean of $\Theta$, and $\overline{\text{Dev}}({\Theta})$ is the posterior mean of $\text{Dev}(\Theta)$. A model with a smaller DIC indicates a preferred model.

Another model comparison criterion is the logarithm of the pseudo-Marginal likelihood
\citep[LPML;][]{ibrahim2013bayesian}, which is obtained
through the conditional predictive ordinate (CPO) values.  With $\bm{y}^*_{(-i)} = (\bm{y}_1,\ldots, \bm{y}_{i-1}, \bm{y}_{i+1},\ldots, \bm{y}_N)$ denoting the binary response vector with the~$i$th subject response deleted, CPO is regarded as leave-one-out-cross-validation under Bayesian framework, and it estimates the probability of observing~$\bm{y}_i$ in the future after having already observed $\bm{y}^{*}_{(-i)}$.
The CPO for the~$i$th subject is calculated as:
\begin{equation}
\label{eq:CPO}
\text{CPO}_i = \int f(\bm{y}_i|\bm{b}, \bm{\theta}) \pi(\bm{b}, \bm{\theta}|\bm{y}^{*}_{(-i)}) d(\bm{b}, \bm{\theta}),
\end{equation} 
where 
\begin{equation*}
\pi(\bm{b}, \bm{\theta}|\bm{y}^{*}_{(-i)}) = \frac{\prod_{q \ne i}
	f(\bm{y}_q|\bm{b}, \bm{\theta} )\pi(\bm{b}, \bm{\theta}| \bm{y}^*_{(-i)})}{c\big(\bm{y}^*_{(-i)}\big)},
\end{equation*}
and $c\big(\bm{y}^*_{(-i)}\big)$ is the normalizing constant. Within the Bayesian framework, a Monte Carlo estimate of the CPO is obtained as:
\begin{equation}
\label{eq:CPOest}
\widehat{\text{CPO}}_i^{-1} = \frac{1}{M} \sum_{t=1}^{M}
\frac{1}{f(\bm{y}_i | \bm{b}_t, \bm{\theta}_t)},
\end{equation}
An estimate of
the LPML subsequently is calculated as:
\begin{equation}
\label{eq:LPML}
\widehat{\text{LPML}} = \sum_{i=1}^{n} \log(\widehat{\text{CPO}}_i).
\end{equation}
A model with a larger LPML value will be selected. 

In addition, the effective number of parameters $p_D$ is applied to compare the Rasch model and the proposed one from a model complexity point of view. A larger number of $p_D$ indicates higher complexity of the model. Besides, the area under the curve (AUC) value is calculated to measure the prediction precision of the model, which is computed via the \textsf{R} package \textbf{pROC} \citep{Rpkg:pROC}. A higher AUC value indicates the better performance under the prediction.

\section{Simulation}\label{sec:simu}
In order to evaluate the performance of the proposed method for clustering and parameter estimation,  simulation studies are conducted in this section. Two situations with different number of items and subjects were considered, and under each situation, two designs were set up.  For both designs, 100 datasets were generated with $N$ subjects and $J$ items in each dataset. The binary response $y_{ij}$ with the probability $p_{ij}$ was generated according to (\ref{eq: Rasch}). Three clusters were assumed for both the subject parameters $\theta_i (i = 1,\cdots,N)$ and the item parameters $b_j (j=1,\cdots, J)$. 
In the first simulation design, the values of $b_j (j = 1, \cdots, J)$ and $\theta_i (i = 1, \cdots, N)$ were both randomly chosen from $\{ -2, 0, 2\}$. Take $b_j$'s for an example. The $j$th difficulty parameter was randomly assigned one of the values of $\{-2,0,2\}$ as its true value, through which, three clusters for the difficulty parameters were formed. In the second simulation design, errors following a normal distribution $N(0, 0.5^2)$ were added to the values of $b_j (j = 1, \cdots, J)$ and $\theta_i (i = 1, \cdots, N)$. Note that the second condition is a more challenging situation because some disturbance of the clustering of items and persons are added in the data by the error $N(0, 0.5^2)$.

The performance of the posterior estimates were evaluated by the mean absolute bias (MAB), the mean standard deviation (MSD) and the mean of mean squared error (MMSE) 
in the
following ways, take $\theta_i$ as an example:
\begin{gather*}
\text{MAB} = \frac{1}{100} \sum_{r=1}^{100}\frac{1}{N} \sum_{i=1}^{N}
\left|\hat{\theta}_{ir} - \theta_{i} \right|, \\
\text{MSD} = 
\sqrt{\frac{1}{100}
	\sum_{r=1}^{100} \frac{1}{N} \sum_{i=1}^{N}\left(\hat{\theta}_{ir} - \bar{\hat{\theta}}_{i}
	\right)^2},\\
\text{MMSE} =  \frac{1}{100}
\sum_{r=1}^{100}
\frac{1}{N} \sum_{i=1}^{N} \left(\hat{\theta}_{ir} - \theta_{i} \right)^2.
\end{gather*}

To evaluate the accuracy of clustering, apart from showing the frequency of cluster numbers in the simulations, the Precision, Recall and Rand Index \citep[RI;][]{rand1971objective} were also applied. 
Let $TP$, $FP$, $TN$ and $FN$ denote the true positive, false positive, true negative and false negative scores, respectively. The Precision was calculated via $TP/(TP+FP)$, the Recall was calculated via $TP/(TP+FN)$, and the RI was obtained by $(TP+TN)/(TP+FP+TN+FN)$ \citep{ma2019bayesianspatial}. 
The average RI (MRI) was calculated as the mean of RIs of the 100 simulations, and similarly the average Precision (MP) and average Recall (MR). For both these three measures, a higher value represents higher accuracy of clustering. 
In addition, the average effective number (MpD), the average AUC (MAUC), and the average computation time in seconds (MTime) were also calculated to compare the complexity of the Bayesian Rasch model and the proposed model. 
Under different settings, with a thinning interval of 2, 10000 samples were kept for calculating the posterior estimates after a burn-in of 20000 samples. Convergence was checked via the traceplots and autocorrelation plots of the chains. 

We firstly consider a situation with $N=200$ subjects and $J=60$ items. 
The MpD, MAUC and MTime values of the two models under two designs are shown in Table \ref{tab:mpd}. 

\begin{table}[h!]
	\centering
	\caption{Model performance results of the proposed model and the Bayesian Rasch model under two designs with $N=200$ and $J=60$}\label{tab:mpd}
	\begin{tabular}{lccccccc}
		\toprule
		\multirow{2}{*}{Design} & \multicolumn{3}{c}{The proposed model } && \multicolumn{3}{c}{Bayesian Rasch model} \\
		\cline{2-4} \cline{6-8}
		& MpD & MAUC & MTime & & MpD & MAUC & MTime\\
		\midrule 
		The first design & 27.698 & 0.831 & 4126.742 & & 243.029 &0.839  & 507.783 \\
		The second design & 85.518 &0.834 & 4225.178& &242.823&0.852 & 515.304\\
		\bottomrule
	\end{tabular}
\end{table} 

From Table \ref{tab:mpd}, we see that under two different designs with $N=200$ and $J=60$, the MpD values of the Bayesian Rasch model are larger than those of the proposed model, meaning that the Bayesian Rasch model has greater complexity and has more parameters to be estimated than the proposed model. On model prediction performance, the MAUC values of the two models are merely the same, which indicates that the prediction precision of the two models under different simulation designs are similar. On the average computation time, the proposed model requires much more time than the Bayesian Rasch model. 

\begin{figure}[hbp] 
	\centering
	\subfigure[Number of clusters of $\theta$ under design 1]{
		\includegraphics[scale=0.3]{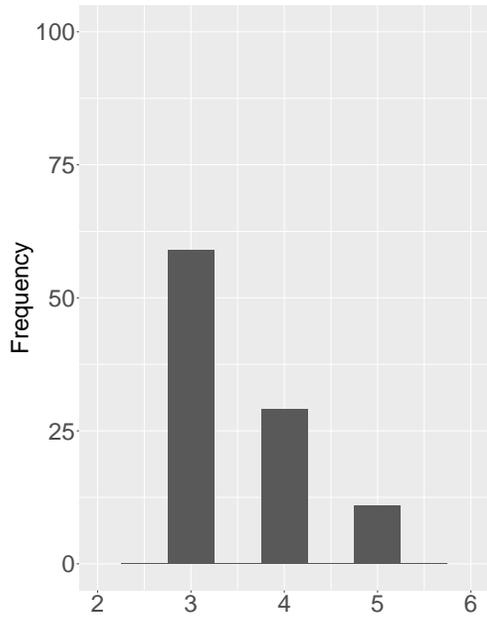}
	}
	\quad
	\subfigure[Number of clusters of $b$ under design 1]{
		\includegraphics[scale=0.3]{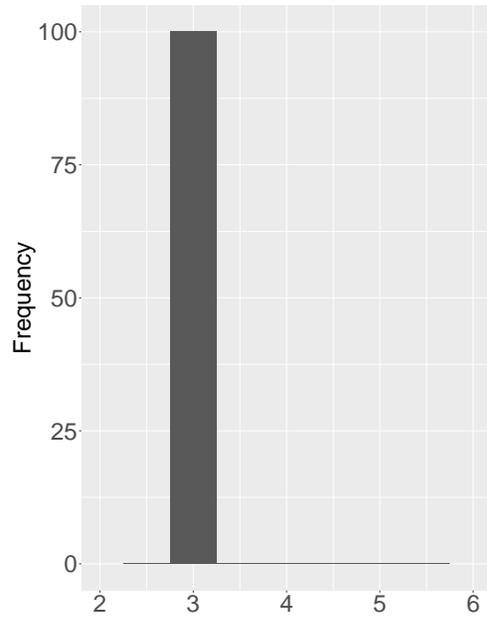}
	}
	\quad
	\subfigure[Number of clusters of $\theta$ under design 2]{
		\includegraphics[scale=0.3]{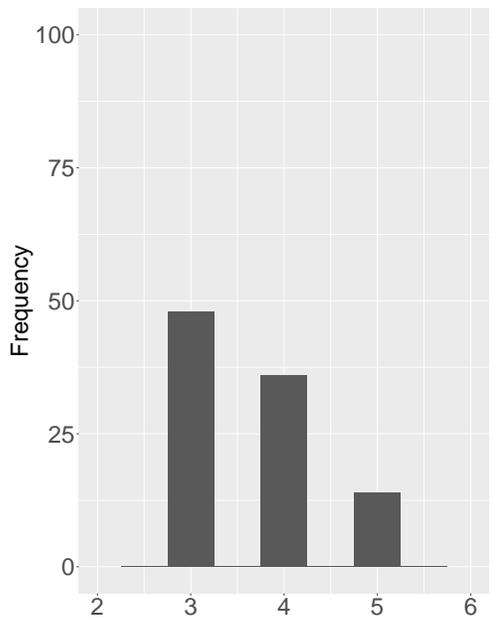}
	}
	\quad
	\subfigure[Number of clusters of $b$ under design 2]{
		\includegraphics[scale=0.3]{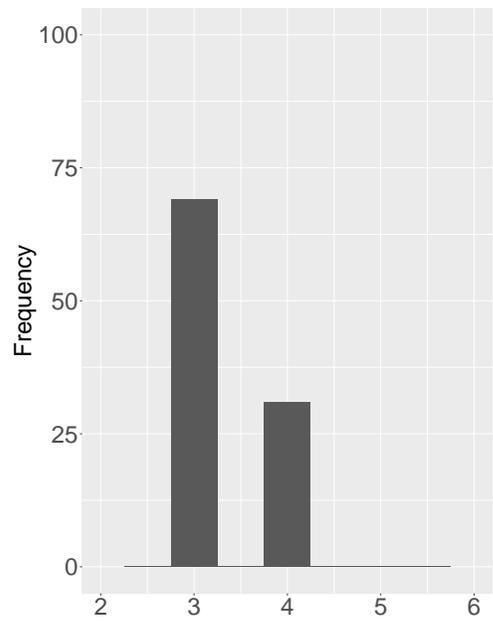}
	}
	\caption{Frequency of number of clusters of $\theta$ and $b$ under different designs with $N=200$ and $J=60$}
	\label{fig:clusterRes}
\end{figure}

The frequencies of the number of clusters for $\bm{\theta}$ and $\bm{b}$ under these two designs with $N=200$ and $J=60$ are shown in Figure~\ref{fig:clusterRes}. Under design 1, the percentages of clustering the $\theta_i$'s and the $b_j$'s into 3 clusters are nearly 60\% and 100\%, respectively. These results mean that for $\theta_i$'s, about 60 out of 100 replicates, the number of clusters is correctly inferenced, and for $b_j$'s, the proportion is 100\%. In addition, the MRI values in Table \ref{tab:simu} show the good performance of clustering allocation, which reach to 0.961 and 1.000 for these parameters. The MRI values indicate that about 96\% and 100\% of the time, any two persons or items that are in the same cluster are correctly grouped together. The MP and MR values are consistent with the MRI values, which verify our conclusions.   
Under design 2, the percentages of correct clustering and both the MRI, MP and MR values of  the two parameters are lower compared to design 1, but the most frequently chosen cluster numbers are also 3 clusters and the MRI values are both larger than 0.85, meaning that when the true values of $\theta_i$'s and $b_j$'s are generated with errors, the precision of clustering are still high.


The results of posterior estimates under two designs with $N=200$ and $J=60$ are also shown in Table \ref{tab:simu}. The MABs and MMSEs of the parameters are both within a reasonable range. In general, performance of posterior estimates under design 1 is better than those under design 2.

\begin{table}[h!]
	\centering
	\caption{Simulation results of the proposed model in two simulation designs with $N=200$ and $J=60$}\label{tab:simu}
	\begin{tabular}{lccccccc}
		\toprule
		Design & Parameter& MAB & MSD &MMSE & MRI & MP & MR\\
		\midrule 
		\multirow{2}{*}{The first design} & $\theta$ &  0.582 & 0.046  & 0.490  & 0.961 & 0.960 & 0.922 \\
		& $b$ & 0.590 & 0.041 & 0.509 & 1.000  & 1.000 & 1.000 \\
		\multirow{2}{*}{The second design}  &$\theta$ &  0.729 & 0.047 & 0.767  & 0.862 & 0.787 & 0.803\\
		& $b$ & 0.723 & 0.060 & 0.778 & 0.899 & 0.862 & 0.833 \\
		\bottomrule
	\end{tabular}
\end{table} 

%

Another situation with $N=200$ subjects and $J=100$ items are considered. Similarly, the MpD, MAUC and MTime values of these two models under two designs are shown in Table \ref{tab:mpd2}. It is shown that the MpD values of the proposed model under two different settings are much lower than those of the Bayesian Rasch model, especially under the first design. The similar MAUC values show that these two models have merely the same prediction precision, and the proposed model still has high precision in addition to clustering. However, the proposed model spends longer time for computation than the Bayesian Rasch model. These conclusions are similar to those when $N=200$ and $J=60$. 

\begin{table}[h!]
	\centering
	\caption{Model performance results of the proposed model and the Bayesian Rasch model under two designs with $N=200$ and $J=100$}\label{tab:mpd2}
	\begin{tabular}{lccccccc}
		\toprule
		\multirow{2}{*}{Design} & \multicolumn{3}{c}{The proposed model } && \multicolumn{3}{c}{Bayesian Rasch model} \\
		\cline{2-4} \cline{6-8}
		& MpD & MAUC & MTime & & MpD & MAUC & MTime\\
		\midrule 
		The first design & 30.819 & 0.832 & 7158.913 && 289.058 & 0.840 & 999.524 \\
		The second design & 90.807 & 0.834 & 7487.103& & 288.721 & 0.853 & 971.540\\
		\bottomrule
	\end{tabular}
\end{table} 

\begin{figure}[htbp] 
	\centering
	\subfigure[Number of clusters of $\theta$ under design 1]{
		\includegraphics[scale=0.3]{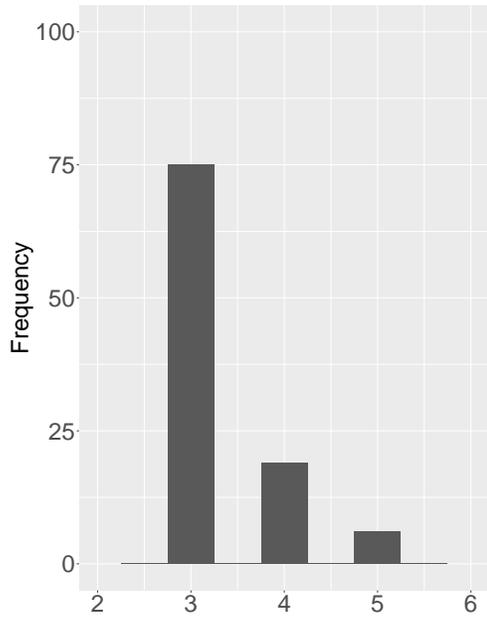}
	}
	\quad
	\subfigure[Number of clusters of $b$ under design 1]{
		\includegraphics[scale=0.3]{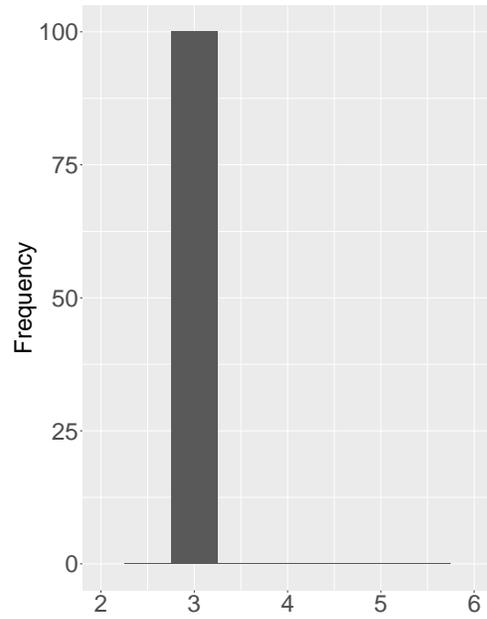}
	}
	\quad
	\subfigure[Number of clusters of $\theta$ under design 2]{
		\includegraphics[scale=0.3]{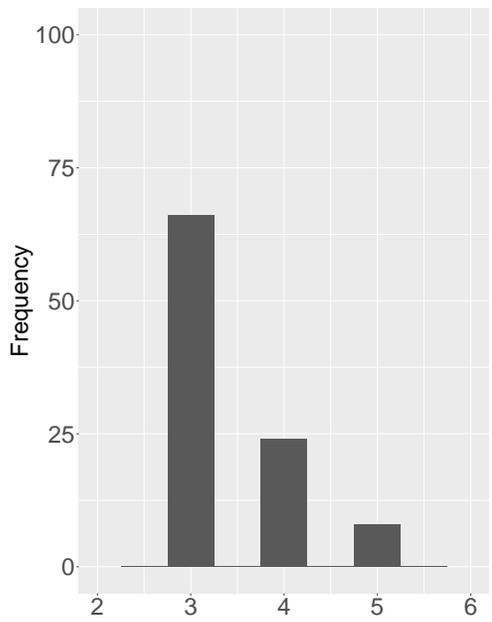}
	}
	\quad
	\subfigure[Number of clusters of $b$ under design 2]{
		\includegraphics[scale=0.3]{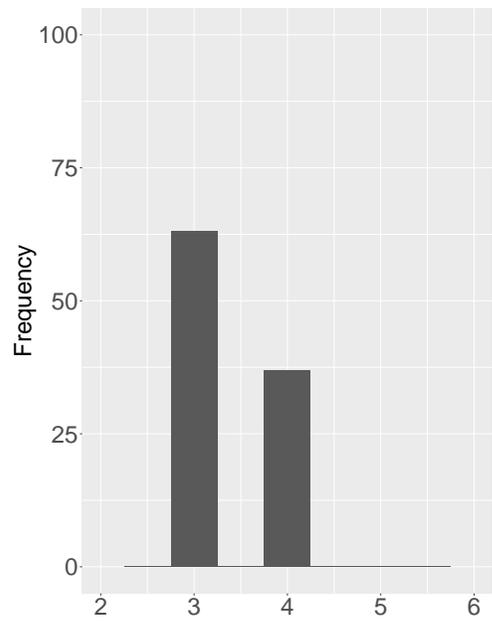}
	}
	\caption{Frequency of number of clusters of $\theta$ and $b$ under different designs with $N=200$ and $J=100$}
	\label{fig:clusterRes2}
\end{figure}

Figure \ref{fig:clusterRes2} shows that frequency of number of clusters under two designs with $N=200$  and $J=100$. Under the first design, nearly 75\% and $100\%$ of the parameters $\theta_i$'s and $b_j$'s are grouped into three clusters. Comparing these results with the results under design 1 with $J =60$, we see that the clustering performance of parameters $\theta_i$'s improves, and the percentage of clustering into 3 clusters increases from nearly 60\% to 75\%. The higher MP, MR and MRI values shown in Table \ref{tab:simu2} also verifies this conclusion. 
Under the second design, the percentages of clustering the parameters $\theta_i$'s and $b_j$'s into 3 clusters are 66\% and 63\%, respectively. Comparing with the results in Figure \ref{fig:clusterRes}, the clustering performance of $\theta_i$'s improves while the clustering performance of $b_j$'s degenerates. This may due to the fact that a greater number of items increase the difficulty of clustering $b_j$'s.  The MRI values in Table \ref{tab:simu2} are both larger than 0.88, meaning that the precision of clustering allocation under this design is high.

In Table \ref{tab:simu2}, the results of posterior estimates under two designs with $N=200$ and $J=100$ are also shown. Comparing with the results in Table \ref{tab:simu}, posterior estimation performance under the first setting with a smaller number of items are relatively better than those under the one with $J=100$.  However, the MABs and MMSEs of the parameters are both within a reasonable range and in general, performance of posterior estimates under design 1 is better than that under design 2. 

\begin{table}[h!]
	\centering
	\caption{Simulation results of the proposed model in two simulation designs with $N=200$ and $J=100$}\label{tab:simu2}
	\begin{tabular}{lccccccc}
		\toprule
		Design & Parameter& MAB & MSD &MMSE & MRI & MP & MR\\
		\midrule 
		\multirow{2}{*}{The first design} & $\theta$ & 0.591 & 0.033 & 0.522 & 0.989 & 0.995 & 0.974\\
		& $b$ & 0.613 & 0.030 & 0.548 & 0.999  & 1.000 & 1.000 \\
		\multirow{2}{*}{The second design}  &$\theta$ & 0.739 & 0.036 & 0.804 & 0.882   & 0.822 & 0.824\\
		& $b$ & 0.741 & 0.046 & 0.817 & 0.897 &0.857 & 0.832\\
		\bottomrule
	\end{tabular}
\end{table} 

\section{Application to English Test Data}\label{sec:real_data}
\subsection{Data Description and Preliminary Analysis} 
\begin{figure} [h!]
	\centering
	\includegraphics[scale = 0.45]{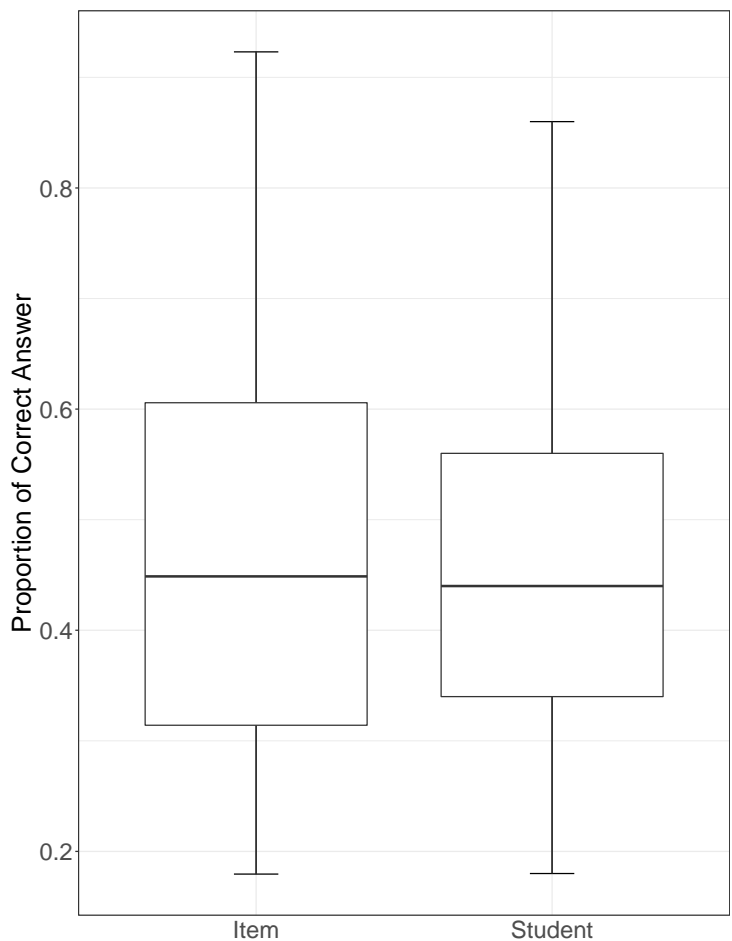}
	\caption{Boxplot of the proportions of correct answer for the items and the students in the English test data.  \label{fig: boxplot}}
\end{figure}

The dataset we analyzed is collected from the English exam results of the 2017-2018 academic year from a public middle school in China. This dataset was previously analyzed in \citet{liu2019comparison}. This exam is a midterm English exam for grade 8 students in fall 2017. In our analysis, the total number of students is 78 and the total number of items is 50. We used this dataset to illustrate the MFM Rasch model estimation.
For each student, the average number of items of correct response is 23.09, while for each item, the average number of students that answer correctly is 36.02, so the average proportion of correct answer is 0.46. The boxplot of the proportions of correct answer for the items and the students are shown in Figure \ref{fig: boxplot}. 

\begin{figure}[h!] 
	\centering  
	\subfigure[Posterior estimates of $\theta$]{
		\includegraphics[scale=0.35]{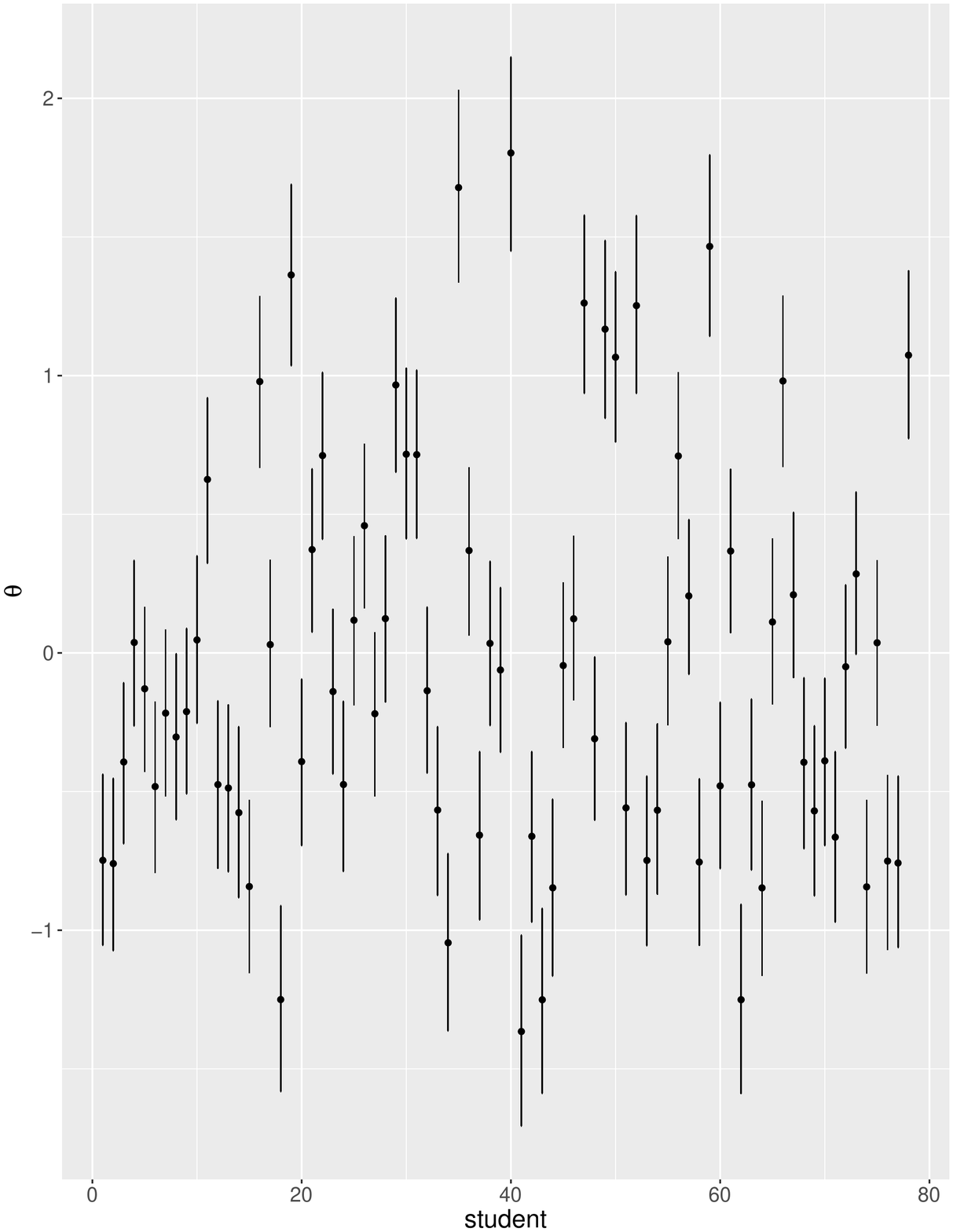}}
	\subfigure[Posterior estimates of $b$]{
		\includegraphics[scale=0.35]{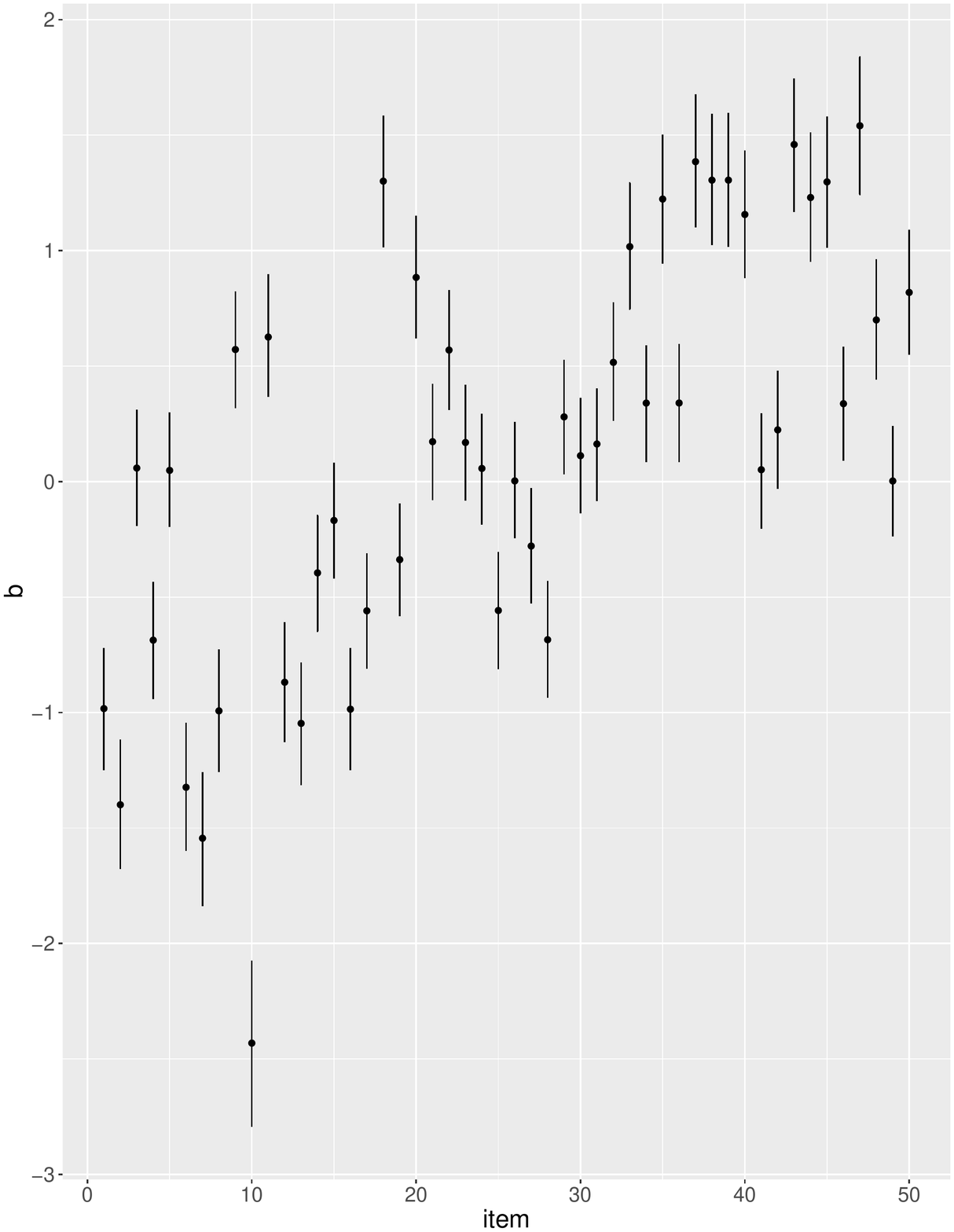}}
	\caption{Posterior estimates and 68\% HPD intervals of parameters in the Rasch model.
	\label{fig:Rasch_res} }
\end{figure}

From Figure \ref{fig: boxplot}, we observe that the correct response proportions vary from different items as well as different students, indicating that the items are of different difficulty levels and the students are of different ability levels. The Bayesian Rasch model in (\ref{eq: Rasch}) was applied to analyze the data. For the unknown parameters $b_j$ and $\theta_i$, the prior distributions were normal distributions, i.e., $b_j \sim N(0,1)$ and $\theta_i \sim N(0,\psi^{-1})$ with hyperprior $\psi \sim \text{Gamma}(0.001, 0.001)$, for $i = 1, \cdots, 78$, $j = 1, \cdots, 50$. 
With a thinning interval of 2, 10000 samples were kept for calculating the posterior estimates after a burn-in of 20000 samples. 
The effective number of parameters $p_D$ of this model was also calculated and turned out to be 107.676. The AUC value was 0.739.
Posterior estimates as well as the 68\% highest posterior density (HPD) intervals of the parameters $b$ and $\theta$ are shown in Figure \ref{fig:Rasch_res}.

From Figure \ref{fig:Rasch_res}, we see that for the items' difficulties and individuals' abilities, the posterior means may be roughly divided into several groups. Take parameter $\theta$'s as an example. Students with the posterior means which are above 0.5 can be grouped together, indicating that the latent abilities of students belonging to this group are close to each other and may be nearly the same.  
For the ones whose posterior means are smaller than 0 may also be grouped together. Due to the patterns we observed, the parameters of different items/individuals may be clustered into different groups. A Rasch model with clustering items and individuals simultaneously would capture the heterogeneity patterns shown in the data.

\subsection{MFM Rasch Clustering Analysis}
In order to detect the clusters within the parameters, the proposed methodology was applied to the English test data. 
 We firstly applied the model comparison criteria to select the most suitable priors for the parameters $\lambda_b$ and $\lambda_{\theta}$. Three priors were compared, including $\text{Gamma}(1,1)$, $\text{Unif}(0, 1)$ and $\text{Log-Normal}(0,1)$. The results of the model selection criteria are shown in Table \ref{tab:LPML}.

\begin{table}[h!]
	\centering
	\caption{DIC and LPML values for different priors
		for $\lambda_b$ and $\lambda_{\theta}$ in the proposed model}\label{tab:LPML}
	\begin{tabular}{lccc}
		\toprule
		& $\text{Gamma}(1,1)$ &$\text{Unif}(0, 1)$ & $\text{Log-Normal}(0,1)$  \\
		\midrule 
		DIC & 4673.480 & 4681.919 & 4675.817 \\
		LPML & -2279.155 & -2282.574 & -2280.213 \\ 
		\bottomrule
	\end{tabular}
\end{table}

From Table \ref{tab:LPML}, we find that the DIC and LPML criteria both prefer the prior $\text{Gamma}(1, 1)$. Thus, we used this prior for $\lambda_b$ and $\lambda_{\theta}$ in our final analysis. 
With this prior, the effective number of parameters was calculated and the value is 57.414. By comparing this value with that of the Bayesian Rasch model (107.676), we see that the complexity of our proposed model is lower and it fits less parameters compared to the Rasch model. In addition, the AUC value of the proposed model is 0.738, which is very close to that of the Bayesian Rasch model (0.739). This indicates that the proposed model can not only cluster the parameters, but also has a similar prediction performance with the Rasch model.

\begin{figure}[h!] 
	\centering  
	\subfigure[Posterior estimates of $\theta$]{
		\includegraphics[scale=0.35]{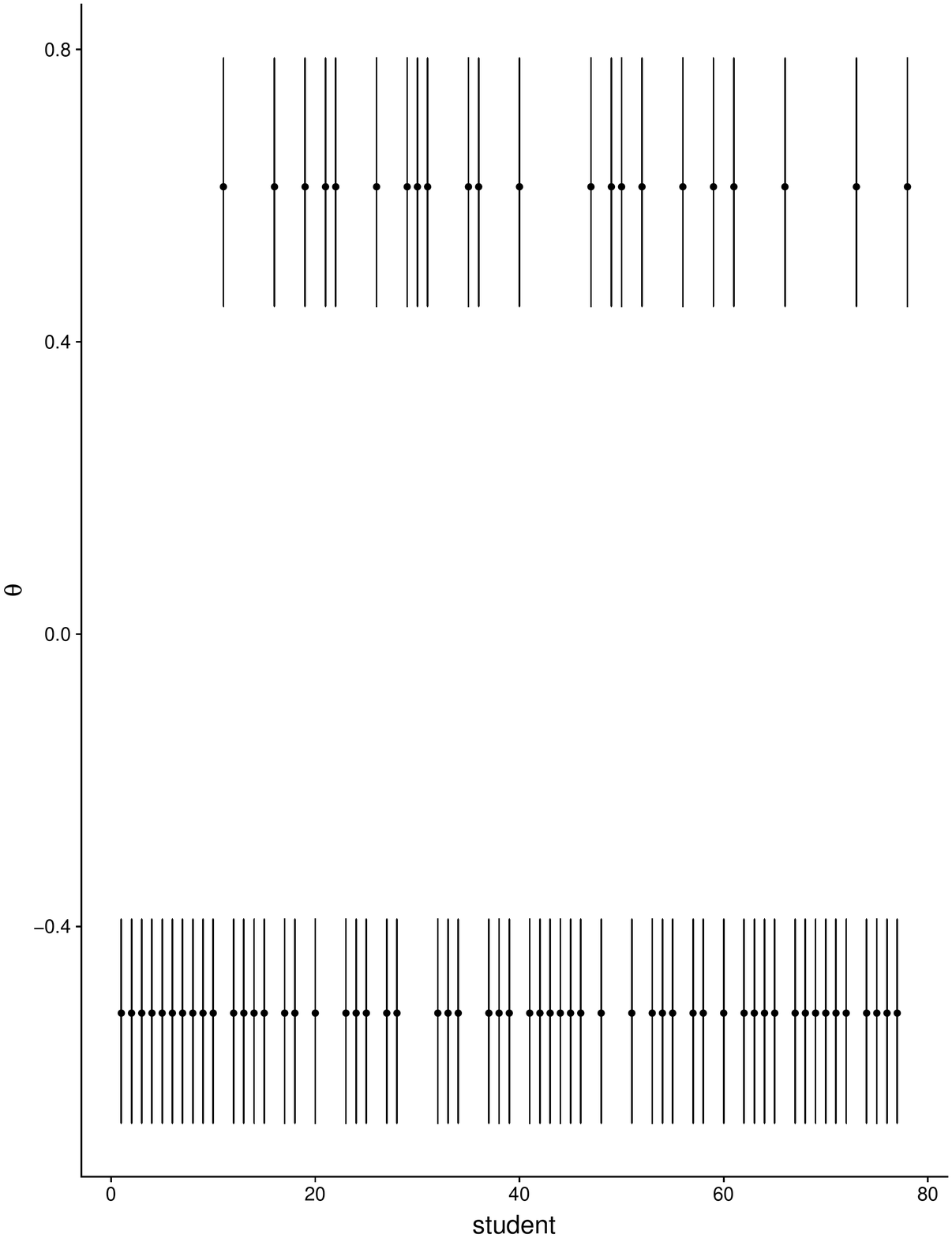}}
	\subfigure[Posterior estimates of $b$]{
		\includegraphics[scale=0.35]{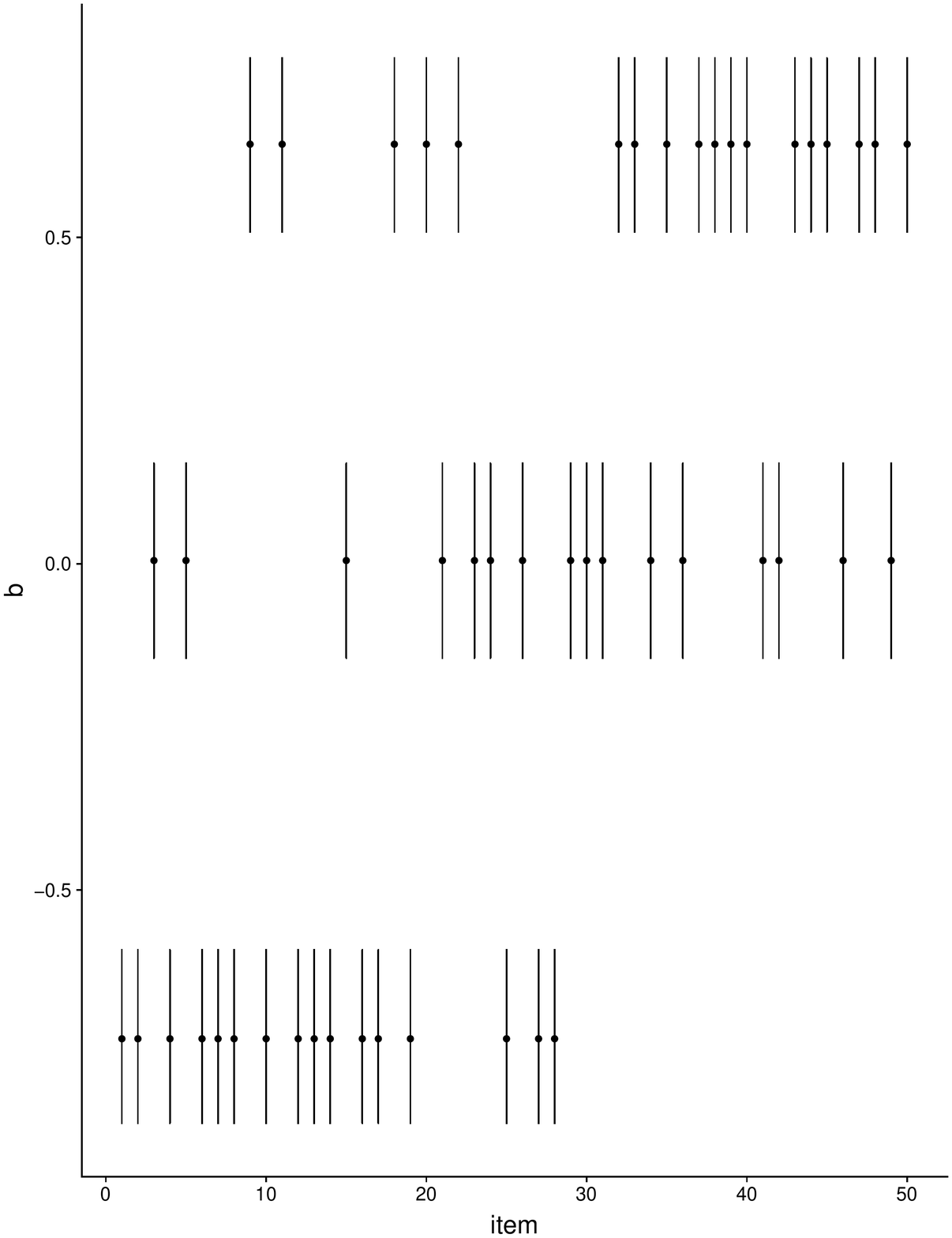}}
	\caption{Clustering allocations of individual abilities and item difficulties in the proposed model
		\label{fig:new_res} }
\end{figure}

Finally, for the latent ability parameters $\theta_i$'s, two clusters are identified, indicating that these 78 students are classified into 2 groups according to different levels of latent ability.  
For the difficulty parameter $b_j$'s, three clusters are identified, meaning that the 50 questions have 3 levels of difficulty. 
The clustering and estimation results are shown in Table \ref{tab:clusterres}. The clustering allocations are shown in Figure \ref{fig:new_res}, where the x-axis represents student index and item number, respectively, in (a) and (b).

\begin{table}[h!]
	\centering
	\caption{The clustering and estimation results for the real data analysis using the proposed model }\label{tab:clusterres}
	\begin{tabular}{lccc}
		\toprule
		\multicolumn{2}{c}{Parameter} & $\bm{\theta}$ &  $\bm{b}$ \\
		\midrule 
	     \multirow{2}{*}{Cluster 1} & Count  & 22 & 18  \\
		 & Estimate & 0.612 (0.448, 0.788) &0.643 (0.508, 0.775)\\
		 \midrule
		 \multirow{2}{*}{Cluster 2} & Count  &56 & 16\\
		& Estimate &  -0.519 (-0.669, -0.390) &0.005 (-0.144, 0.154) \\
		\midrule
		 \multirow{2}{*}{Cluster 3} & Count  &- & 16 \\
		& Estimate &-&  -0.728 (-0.857, -0.592) \\
		\bottomrule
	\end{tabular}
\end{table} 

From Table \ref{tab:clusterres} and Figure \ref{fig:new_res}, we see that for latent ability, students are clustered into 2 groups, and 56 among the 78 students  (71.795\% of the students) are in Cluster 2, whose latent abilities are relatively lower. For the students in Cluster 1, their latent abilities are higher. In order to decide if these results are consistent to the original data, we calculated the average proportion of correct answer of students within these two clusters respectively. The results are shown in Table \ref{tab:ratio}.

\begin{table}[h!]
	\centering
	\caption{Average proportion of correct answer in different clusters of latent ability}\label{tab:ratio}
	\begin{tabular}{lccc}
		\toprule
		Cluster  &  Cluster 1 & Cluster 2 & All \\
		\midrule 
		Proportion of correct answer & 68.45\% & 37.43\% & 46.18\%  \\
		\bottomrule
	\end{tabular}
\end{table} 

From Table \ref{tab:ratio}, we notice that students in Cluster 1 have a proportion of correct answer that is above the average level, while the ones in Cluster 2 have a lower proportion of correct answer, which is consistent with the clustering results of latent ability. 

For item difficulties, three clusters are identified. There are 18 items (36\% of the test length) are grouped into Cluster 1, while for Cluster 2 and Cluster 3, there are 16 items (32\% of the test length) in each cluster. It is shown that questions belonging to Cluster 1 are the most difficult, while the ones in Cluster 3 are the easiest. Similarly, we calculated the proportion of correctly answer the questions in different clusters to see whether the estimation results are consistent to the real data. The average proportions of correct answer in three clusters of items are shown in Table \ref{tab:ratio2}.

\begin{table}[h!]
	\centering
	\caption{Average proportion of correct answer in different clusters of difficulty}\label{tab:ratio2}
	\begin{tabular}{lcccc}
		\toprule
		Cluster  &  Cluster 1 & Cluster 2 & Cluster 3  &  All \\
		\midrule 
		Proportion of correct answer & 26.64\% & 45.59\% &  68.75\%& 46.18\%\\
		\bottomrule
	\end{tabular}
\end{table} 

From Table \ref{tab:ratio2} we see that students have the highest proportion of correct answer to items in Cluster 3, but the lowest proportion in Cluster 1, indicating that items in Cluster 1 are the most difficult while items in Cluster 3 are relatively easier. The difficulty level of items in Cluster 2 are at an average level since the proportion of correct answer in Cluster 2 is similar to the average proportion of correct answer of all the items. These results are also consistent with the conclusions we made according to the clustering and estimation results.

\begin{figure} [h!]
	\centering
	\includegraphics[scale = 0.4]{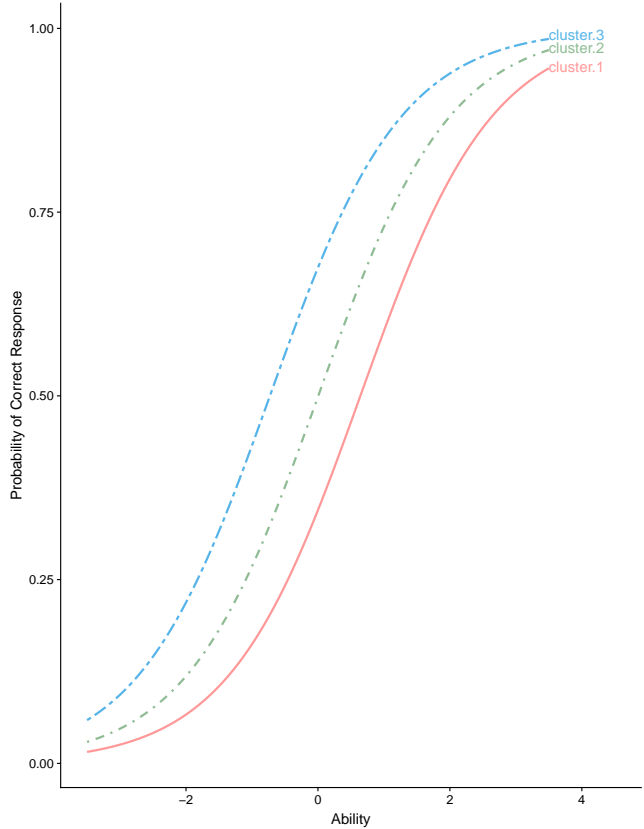}
	\caption{Item characteristic curves of the hierarchical Rasch model with MFM \label{fig:ICC}}
\end{figure}

The item characteristic curves using the item parameter estimates of the three clusters from the proposed model are shown in Figure \ref{fig:ICC}.
Since the item difficulty parameters are clustered into three groups, we have three item characteristic curves for the proposed IRT model. For the three curves with different item difficulty levels, the locations of the curves along the ability scale are different, meaning that the difficulty of the items within different groups vary. The curve of items within cluster 3 is at the left-hand side, representing items in this cluster are the easiest since the probability of correct response is high for low-ability students and approaches 1 for high-ability students. Curve of cluster 2 is in the middle, meaning that items within this cluster are of medium difficulty. The curve for items in cluster 1 is the right-hand curve, representing the items in this cluster are hard items. 

\section{Summary and Discussion}\label{sec:discussion}

In the regular Rasch model, items are treated as invariant across test-takers, thus an item's difficulty is the same across all test-takers. Also, test-takers are assumed to be from a homogeneous group. This homogeneity of item and person parameters is not always feasible in reality. LCA or mixture IRT model has been a major tool to investigate the heterogeneity or the violation of the homogeneity assumption of item and person parameters. LCA could detect heterogeneity in persons, permitting person clustering. Mixture IRT modeling allows person and item heterogeneity, detecting different subgroups of test-takers which have different constellations of item parameters. As pointed out, these LCA and mixture IRT modeling requires a pre-specification of the number of clusters by users before estimation. Also, a two-stage procedure, where several models are estimated and compared to pick the most appropriate model, is not efficient and disregards the uncertainties which take place in the selection procedure. 

In this paper, we proposed a Bayesian nonparametric model for capturing the heterogeneity of items and individuals simultaneously under the Rasch model. Our proposed model, MFM Rasch model, simultaneously estimate the number of clusters and clustering configurations for both items and individuals. It does not require \textit{a priori} specification of the number of clusters by users. Because of the simultaneous clustering of both the item and person parameters, it avoids the problem of ignoring the uncertainties in the two-stage procedure used in LA and mixture IRT modeling. A practical side of the MFM Rasch modeling could include situations where users of the model are interested in identifying clusters of items and persons, so that the information of clusters may be used for their assessment purposes. For example, information on item clusters may be useful in the test construction stage and information on person clusters may also be utilized for student-placement purposes or diagnostic purposes.

Our simulation results and real data analysis indicated that the MFM Rasch estimation could achieve both accurate estimation and clustering performance although the scope of the simulation is limited and the real data set may not be the most typical application scenario regarding the sample size and test length. Future endeavor should include more systematic studies through wider scope of simulations (e.g., varying the number of clusters, sample sizes, and test lengths) to provide a complete picture of the performance of the MFM Rasch model. In addition, 
at least three topics beyond the scope of this paper could be worth further investigation. First, extending the MFM Rasch model into the 2-parameter or 3-parameter item response theory model seems to be an  interesting future work. Second, our proposed computing algorithm is based on stick-breaking construction. Incorporating  P\'{o}lya urn scheme in posterior sample \citep{neal2000markov} will speed up the MCMC algorithm. Third, extending the current model to consider auxiliary information of individuals such as previous exam performance for clustering detection may be pursued in future research. 

%
%


\bibliographystyle{chicago}
\bibliography{irt}

\begin{thebibliography}{}

\bibitem [\protect \citeauthoryear {%
Akaike%
}{%
Akaike%
}{%
{\protect \APACyear {1998}}%
}]{%
akaike1998information}
\APACinsertmetastar {%
akaike1998information}%
\begin{APACrefauthors}%
Akaike, H.%
\end{APACrefauthors}%
\unskip\
\newblock
\APACrefYearMonthDay{1998}{}{}.
\newblock
{\BBOQ}\APACrefatitle {Information theory and an extension of the maximum
  likelihood principle} {Information theory and an extension of the maximum
  likelihood principle}.{\BBCQ}
\newblock
\BIn{} \APACrefbtitle {Selected papers of hirotugu akaike} {Selected papers of
  hirotugu akaike}\ (\BPGS\ 199--213).
\newblock
\APACaddressPublisher{}{Springer}.
\PrintBackRefs{\CurrentBib}

\bibitem [\protect \citeauthoryear {%
Bartolucci%
}{%
Bartolucci%
}{%
{\protect \APACyear {2007}}%
}]{%
bartolucci2007class}
\APACinsertmetastar {%
bartolucci2007class}%
\begin{APACrefauthors}%
Bartolucci, F.%
\end{APACrefauthors}%
\unskip\
\newblock
\APACrefYearMonthDay{2007}{}{}.
\newblock
{\BBOQ}\APACrefatitle {A class of multidimensional IRT models for testing
  unidimensionality and clustering items} {A class of multidimensional irt
  models for testing unidimensionality and clustering items}.{\BBCQ}
\newblock
\APACjournalVolNumPages{Psychometrika}{72}{2}{141}.
\PrintBackRefs{\CurrentBib}

\bibitem [\protect \citeauthoryear {%
Bartolucci%
, Farcomeni%
\BCBL {}\ \BBA {} Scaccia%
}{%
Bartolucci%
\ \protect \BOthers {.}}{%
{\protect \APACyear {2017}}%
}]{%
bartolucci2017nonparametric}
\APACinsertmetastar {%
bartolucci2017nonparametric}%
\begin{APACrefauthors}%
Bartolucci, F.%
, Farcomeni, A.%
\BCBL {}\ \BBA {} Scaccia, L.%
\end{APACrefauthors}%
\unskip\
\newblock
\APACrefYearMonthDay{2017}{}{}.
\newblock
{\BBOQ}\APACrefatitle {A nonparametric multidimensional latent class IRT model
  in a Bayesian framework} {A nonparametric multidimensional latent class irt
  model in a bayesian framework}.{\BBCQ}
\newblock
\APACjournalVolNumPages{Psychometrika}{82}{4}{952--978}.
\PrintBackRefs{\CurrentBib}

\bibitem [\protect \citeauthoryear {%
de Valpine%
\ \protect \BOthers {.}}{%
de Valpine%
\ \protect \BOthers {.}}{%
{\protect \APACyear {2017}}%
}]{%
de2017programming}
\APACinsertmetastar {%
de2017programming}%
\begin{APACrefauthors}%
de Valpine, P.%
, Turek, D.%
, Paciorek, C\BPBI J.%
, Anderson-Bergman, C.%
, Lang, D\BPBI T.%
\BCBL {}\ \BBA {} Bodik, R.%
\end{APACrefauthors}%
\unskip\
\newblock
\APACrefYearMonthDay{2017}{}{}.
\newblock
{\BBOQ}\APACrefatitle {Programming with models: writing statistical algorithms
  for general model structures with {NIMBLE}} {Programming with models: writing
  statistical algorithms for general model structures with {NIMBLE}}.{\BBCQ}
\newblock
\APACjournalVolNumPages{Journal of Computational and Graphical
  Statistics}{26}{2}{403--413}.
\PrintBackRefs{\CurrentBib}

\bibitem [\protect \citeauthoryear {%
Ferguson%
}{%
Ferguson%
}{%
{\protect \APACyear {1973}}%
}]{%
ferguson1973bayesian}
\APACinsertmetastar {%
ferguson1973bayesian}%
\begin{APACrefauthors}%
Ferguson, T\BPBI S.%
\end{APACrefauthors}%
\unskip\
\newblock
\APACrefYearMonthDay{1973}{}{}.
\newblock
{\BBOQ}\APACrefatitle {A {B}ayesian Analysis of Some Nonparametric Problems} {A
  {B}ayesian analysis of some nonparametric problems}.{\BBCQ}
\newblock
\APACjournalVolNumPages{Annals of Statistics}{1}{2}{209-230}.
\PrintBackRefs{\CurrentBib}

\bibitem [\protect \citeauthoryear {%
Geng%
, Bhattacharya%
\BCBL {}\ \BBA {} Pati%
}{%
Geng%
, Bhattacharya%
\BCBL {}\ \BBA {} Pati%
}{%
{\protect \APACyear {2019}}%
}]{%
geng2019probabilistic}
\APACinsertmetastar {%
geng2019probabilistic}%
\begin{APACrefauthors}%
Geng, J.%
, Bhattacharya, A.%
\BCBL {}\ \BBA {} Pati, D.%
\end{APACrefauthors}%
\unskip\
\newblock
\APACrefYearMonthDay{2019}{}{}.
\newblock
{\BBOQ}\APACrefatitle {Probabilistic community detection with unknown number of
  communities} {Probabilistic community detection with unknown number of
  communities}.{\BBCQ}
\newblock
\APACjournalVolNumPages{Journal of the American Statistical
  Association}{114}{526}{893--905}.
\PrintBackRefs{\CurrentBib}

\bibitem [\protect \citeauthoryear {%
Geng%
, Shi%
\BCBL {}\ \BBA {} Hu%
}{%
Geng%
, Shi%
\BCBL {}\ \BBA {} Hu%
}{%
{\protect \APACyear {2019}}%
}]{%
geng2019bayesian}
\APACinsertmetastar {%
geng2019bayesian}%
\begin{APACrefauthors}%
Geng, J.%
, Shi, W.%
\BCBL {}\ \BBA {} Hu, G.%
\end{APACrefauthors}%
\unskip\
\newblock
\APACrefYearMonthDay{2019}{}{}.
\newblock
{\BBOQ}\APACrefatitle {Bayesian Nonparametric Nonhomogeneous Poisson Process
  with Applications to USGS Earthquake Data} {Bayesian nonparametric
  nonhomogeneous poisson process with applications to usgs earthquake
  data}.{\BBCQ}
\newblock
\APACjournalVolNumPages{arXiv preprint arXiv:1907.03186}{}{}{}.
\PrintBackRefs{\CurrentBib}

\bibitem [\protect \citeauthoryear {%
Gnaldi%
, Bacci%
\BCBL {}\ \BBA {} Bartolucci%
}{%
Gnaldi%
\ \protect \BOthers {.}}{%
{\protect \APACyear {2016}}%
}]{%
gnaldi2016multilevel}
\APACinsertmetastar {%
gnaldi2016multilevel}%
\begin{APACrefauthors}%
Gnaldi, M.%
, Bacci, S.%
\BCBL {}\ \BBA {} Bartolucci, F.%
\end{APACrefauthors}%
\unskip\
\newblock
\APACrefYearMonthDay{2016}{}{}.
\newblock
{\BBOQ}\APACrefatitle {A multilevel finite mixture item response model to
  cluster examinees and schools} {A multilevel finite mixture item response
  model to cluster examinees and schools}.{\BBCQ}
\newblock
\APACjournalVolNumPages{Advances in Data Analysis and
  Classification}{10}{1}{53--70}.
\PrintBackRefs{\CurrentBib}

\bibitem [\protect \citeauthoryear {%
Hagenaars%
\ \BBA {} McCutcheon%
}{%
Hagenaars%
\ \BBA {} McCutcheon%
}{%
{\protect \APACyear {2002}}%
}]{%
hagenaars2002applied}
\APACinsertmetastar {%
hagenaars2002applied}%
\begin{APACrefauthors}%
Hagenaars, J\BPBI A.%
\BCBT {}\ \BBA {} McCutcheon, A\BPBI L.%
\end{APACrefauthors}%
\unskip\
\newblock
\APACrefYear{2002}.
\newblock
\APACrefbtitle {Applied latent class analysis} {Applied latent class analysis}.
\newblock
\APACaddressPublisher{}{Cambridge University Press}.
\PrintBackRefs{\CurrentBib}

\bibitem [\protect \citeauthoryear {%
Hu%
, Geng%
, Xue%
\BCBL {}\ \BBA {} Sang%
}{%
Hu%
\ \protect \BOthers {.}}{%
{\protect \APACyear {2020}}%
}]{%
hu2020bayesian}
\APACinsertmetastar {%
hu2020bayesian}%
\begin{APACrefauthors}%
Hu, G.%
, Geng, J.%
, Xue, Y.%
\BCBL {}\ \BBA {} Sang, H.%
\end{APACrefauthors}%
\unskip\
\newblock
\APACrefYearMonthDay{2020}{}{}.
\newblock
{\BBOQ}\APACrefatitle {Bayesian Spatial Homogeneity Pursuit of Functional Data:
  an Application to the US Income Distribution} {Bayesian spatial homogeneity
  pursuit of functional data: an application to the us income
  distribution}.{\BBCQ}
\newblock
\APACjournalVolNumPages{arXiv preprint arXiv:2002.06663}{}{}{}.
\PrintBackRefs{\CurrentBib}

\bibitem [\protect \citeauthoryear {%
Ibrahim%
, Chen%
\BCBL {}\ \BBA {} Sinha%
}{%
Ibrahim%
\ \protect \BOthers {.}}{%
{\protect \APACyear {2013}}%
}]{%
ibrahim2013bayesian}
\APACinsertmetastar {%
ibrahim2013bayesian}%
\begin{APACrefauthors}%
Ibrahim, J\BPBI G.%
, Chen, M\BHBI H.%
\BCBL {}\ \BBA {} Sinha, D.%
\end{APACrefauthors}%
\unskip\
\newblock
\APACrefYear{2013}.
\newblock
\APACrefbtitle {Bayesian survival analysis} {Bayesian survival analysis}.
\newblock
\APACaddressPublisher{}{Springer Science \& Business Media}.
\PrintBackRefs{\CurrentBib}

\bibitem [\protect \citeauthoryear {%
Ishwaran%
\ \BBA {} James%
}{%
Ishwaran%
\ \BBA {} James%
}{%
{\protect \APACyear {2001}}%
}]{%
ishwaran2001gibbs}
\APACinsertmetastar {%
ishwaran2001gibbs}%
\begin{APACrefauthors}%
Ishwaran, H.%
\BCBT {}\ \BBA {} James, L\BPBI F.%
\end{APACrefauthors}%
\unskip\
\newblock
\APACrefYearMonthDay{2001}{}{}.
\newblock
{\BBOQ}\APACrefatitle {Gibbs sampling methods for stick-breaking priors} {Gibbs
  sampling methods for stick-breaking priors}.{\BBCQ}
\newblock
\APACjournalVolNumPages{Journal of the American Statistical
  Association}{96}{453}{161--173}.
\PrintBackRefs{\CurrentBib}

\bibitem [\protect \citeauthoryear {%
Lazarsfeld%
\ \BBA {} Henry%
}{%
Lazarsfeld%
\ \BBA {} Henry%
}{%
{\protect \APACyear {1968}}%
}]{%
lazarsfeld1968latent}
\APACinsertmetastar {%
lazarsfeld1968latent}%
\begin{APACrefauthors}%
Lazarsfeld, P\BPBI F.%
\BCBT {}\ \BBA {} Henry, N\BPBI W.%
\end{APACrefauthors}%
\unskip\
\newblock
\APACrefYear{1968}.
\newblock
\APACrefbtitle {Latent structure analysis} {Latent structure analysis}.
\newblock
\APACaddressPublisher{}{Houghton Mifflin Co.}
\PrintBackRefs{\CurrentBib}

\bibitem [\protect \citeauthoryear {%
Liu%
}{%
X.~Liu%
}{%
{\protect \APACyear {2019}}%
}]{%
liu2019three}
\APACinsertmetastar {%
liu2019three}%
\begin{APACrefauthors}%
Liu, X.%
\end{APACrefauthors}%
\unskip\
\newblock
\APACrefYear{2019}.
\unskip\
\newblock
\APACrefbtitle {Three Contributions to Latent Variable Modeling} {Three
  contributions to latent variable modeling}\ \APACtypeAddressSchool
  {\BUPhD}{}{}.
\unskip\
\newblock
\APACaddressSchool {}{Columbia University}.
\PrintBackRefs{\CurrentBib}

\bibitem [\protect \citeauthoryear {%
Y.~Liu%
, Hu%
, Cao%
, Wang%
\BCBL {}\ \BBA {} Chen%
}{%
Y.~Liu%
\ \protect \BOthers {.}}{%
{\protect \APACyear {2019}}%
}]{%
liu2019comparison}
\APACinsertmetastar {%
liu2019comparison}%
\begin{APACrefauthors}%
Liu, Y.%
, Hu, G.%
, Cao, L.%
, Wang, X.%
\BCBL {}\ \BBA {} Chen, M\BHBI H.%
\end{APACrefauthors}%
\unskip\
\newblock
\APACrefYearMonthDay{2019}{}{}.
\newblock
{\BBOQ}\APACrefatitle {A comparison of Monte Carlo methods for computing
  marginal likelihoods of item response theory models} {A comparison of monte
  carlo methods for computing marginal likelihoods of item response theory
  models}.{\BBCQ}
\newblock
\APACjournalVolNumPages{Journal of the Korean Statistical Society}{}{}{}.
\PrintBackRefs{\CurrentBib}

\bibitem [\protect \citeauthoryear {%
Lunn%
, Thomas%
, Best%
\BCBL {}\ \BBA {} Spiegelhalter%
}{%
Lunn%
\ \protect \BOthers {.}}{%
{\protect \APACyear {2000}}%
}]{%
lunn2000winbugs}
\APACinsertmetastar {%
lunn2000winbugs}%
\begin{APACrefauthors}%
Lunn, D\BPBI J.%
, Thomas, A.%
, Best, N.%
\BCBL {}\ \BBA {} Spiegelhalter, D.%
\end{APACrefauthors}%
\unskip\
\newblock
\APACrefYearMonthDay{2000}{}{}.
\newblock
{\BBOQ}\APACrefatitle {WinBUGS-a Bayesian modelling framework: concepts,
  structure, and extensibility} {Winbugs-a bayesian modelling framework:
  concepts, structure, and extensibility}.{\BBCQ}
\newblock
\APACjournalVolNumPages{Statistics and computing}{10}{4}{325--337}.
\PrintBackRefs{\CurrentBib}

\bibitem [\protect \citeauthoryear {%
Ma%
, Xue%
\BCBL {}\ \BBA {} Hu%
}{%
Ma%
\ \protect \BOthers {.}}{%
{\protect \APACyear {2019}}%
}]{%
ma2019bayesianspatial}
\APACinsertmetastar {%
ma2019bayesianspatial}%
\begin{APACrefauthors}%
Ma, Z.%
, Xue, Y.%
\BCBL {}\ \BBA {} Hu, G.%
\end{APACrefauthors}%
\unskip\
\newblock
\APACrefYearMonthDay{2019}{}{}.
\newblock
{\BBOQ}\APACrefatitle {Heterogeneous Regression Models for Clusters of Spatial
  Dependent Data} {Heterogeneous regression models for clusters of spatial
  dependent data}.{\BBCQ}
\newblock
\APACjournalVolNumPages{}{}{}{arXiv:1907.02212}.
\PrintBackRefs{\CurrentBib}

\bibitem [\protect \citeauthoryear {%
Miller%
\ \BBA {} Harrison%
}{%
Miller%
\ \BBA {} Harrison%
}{%
{\protect \APACyear {2013}}%
}]{%
miller2013simple}
\APACinsertmetastar {%
miller2013simple}%
\begin{APACrefauthors}%
Miller, J\BPBI W.%
\BCBT {}\ \BBA {} Harrison, M\BPBI T.%
\end{APACrefauthors}%
\unskip\
\newblock
\APACrefYearMonthDay{2013}{}{}.
\newblock
{\BBOQ}\APACrefatitle {A simple example of {D}irichlet process mixture
  inconsistency for the number of components} {A simple example of {D}irichlet
  process mixture inconsistency for the number of components}.{\BBCQ}
\newblock
\BIn{} \APACrefbtitle {Advances in Neural Information Processing Systems}
  {Advances in neural information processing systems}\ (\BPGS\ 199--206).
\PrintBackRefs{\CurrentBib}

\bibitem [\protect \citeauthoryear {%
Miller%
\ \BBA {} Harrison%
}{%
Miller%
\ \BBA {} Harrison%
}{%
{\protect \APACyear {2018}}%
}]{%
miller2018mixture}
\APACinsertmetastar {%
miller2018mixture}%
\begin{APACrefauthors}%
Miller, J\BPBI W.%
\BCBT {}\ \BBA {} Harrison, M\BPBI T.%
\end{APACrefauthors}%
\unskip\
\newblock
\APACrefYearMonthDay{2018}{}{}.
\newblock
{\BBOQ}\APACrefatitle {Mixture models with a prior on the number of components}
  {Mixture models with a prior on the number of components}.{\BBCQ}
\newblock
\APACjournalVolNumPages{Journal of the American Statistical
  Association}{113}{521}{340--356}.
\PrintBackRefs{\CurrentBib}

\bibitem [\protect \citeauthoryear {%
Miyazaki%
\ \BBA {} Hoshino%
}{%
Miyazaki%
\ \BBA {} Hoshino%
}{%
{\protect \APACyear {2009}}%
}]{%
miyazaki2009bayesian}
\APACinsertmetastar {%
miyazaki2009bayesian}%
\begin{APACrefauthors}%
Miyazaki, K.%
\BCBT {}\ \BBA {} Hoshino, T.%
\end{APACrefauthors}%
\unskip\
\newblock
\APACrefYearMonthDay{2009}{}{}.
\newblock
{\BBOQ}\APACrefatitle {A {B}ayesian semiparametric item response model with
  Dirichlet process priors} {A {B}ayesian semiparametric item response model
  with dirichlet process priors}.{\BBCQ}
\newblock
\APACjournalVolNumPages{Psychometrika}{74}{3}{375--393}.
\PrintBackRefs{\CurrentBib}

\bibitem [\protect \citeauthoryear {%
Neal%
}{%
Neal%
}{%
{\protect \APACyear {2000}}%
}]{%
neal2000markov}
\APACinsertmetastar {%
neal2000markov}%
\begin{APACrefauthors}%
Neal, R\BPBI M.%
\end{APACrefauthors}%
\unskip\
\newblock
\APACrefYearMonthDay{2000}{}{}.
\newblock
{\BBOQ}\APACrefatitle {Markov chain sampling methods for {D}irichlet process
  mixture models} {Markov chain sampling methods for {D}irichlet process
  mixture models}.{\BBCQ}
\newblock
\APACjournalVolNumPages{Journal of Computational and Graphical
  Statistics}{9}{2}{249--265}.
\PrintBackRefs{\CurrentBib}

\bibitem [\protect \citeauthoryear {%
Pan%
\ \BBA {} Huang%
}{%
Pan%
\ \BBA {} Huang%
}{%
{\protect \APACyear {2014}}%
}]{%
pan2014bayesian}
\APACinsertmetastar {%
pan2014bayesian}%
\begin{APACrefauthors}%
Pan, J\BHBI C.%
\BCBT {}\ \BBA {} Huang, G\BHBI H.%
\end{APACrefauthors}%
\unskip\
\newblock
\APACrefYearMonthDay{2014}{}{}.
\newblock
{\BBOQ}\APACrefatitle {Bayesian inferences of latent class models with an
  unknown number of classes} {Bayesian inferences of latent class models with
  an unknown number of classes}.{\BBCQ}
\newblock
\APACjournalVolNumPages{Psychometrika}{79}{4}{621--646}.
\PrintBackRefs{\CurrentBib}

\bibitem [\protect \citeauthoryear {%
Pitman%
}{%
Pitman%
}{%
{\protect \APACyear {1995}}%
}]{%
pitman1995exchangeable}
\APACinsertmetastar {%
pitman1995exchangeable}%
\begin{APACrefauthors}%
Pitman, J.%
\end{APACrefauthors}%
\unskip\
\newblock
\APACrefYearMonthDay{1995}{}{}.
\newblock
{\BBOQ}\APACrefatitle {Exchangeable and partially exchangeable random
  partitions} {Exchangeable and partially exchangeable random
  partitions}.{\BBCQ}
\newblock
\APACjournalVolNumPages{Probability Theory and Related
  Fields}{102}{2}{145--158}.
\PrintBackRefs{\CurrentBib}

\bibitem [\protect \citeauthoryear {%
Plummer%
\ \protect \BOthers {.}}{%
Plummer%
\ \protect \BOthers {.}}{%
{\protect \APACyear {2003}}%
}]{%
plummer2003jags}
\APACinsertmetastar {%
plummer2003jags}%
\begin{APACrefauthors}%
Plummer, M.%
\BCBT {}\ \BOthersPeriod {.}
\end{APACrefauthors}%
\unskip\
\newblock
\APACrefYearMonthDay{2003}{}{}.
\newblock
{\BBOQ}\APACrefatitle {JAGS: A program for analysis of Bayesian graphical
  models using Gibbs sampling} {Jags: A program for analysis of bayesian
  graphical models using gibbs sampling}.{\BBCQ}
\newblock
\BIn{} \APACrefbtitle {Proceedings of the 3rd international workshop on
  distributed statistical computing} {Proceedings of the 3rd international
  workshop on distributed statistical computing}\ (\BVOL~124, \BPG~10).
\PrintBackRefs{\CurrentBib}

\bibitem [\protect \citeauthoryear {%
{R Core Team}%
}{%
{R Core Team}%
}{%
{\protect \APACyear {2013}}%
}]{%
Rlanguage2013}
\APACinsertmetastar {%
Rlanguage2013}%
\begin{APACrefauthors}%
{R Core Team}.%
\end{APACrefauthors}%
\unskip\
\newblock
\APACrefYearMonthDay{2013}{}{}.
\newblock
{\BBOQ}\APACrefatitle {R: A Language and Environment for Statistical Computing}
  {R: A language and environment for statistical computing}{\BBCQ}\
  [\bibcomputersoftwaremanual].
\newblock
\APACaddressPublisher{Vienna, Austria}{}.
\newblock
\begin{APACrefURL} \url{http://www.R-project.org/} \end{APACrefURL}
\PrintBackRefs{\CurrentBib}

\bibitem [\protect \citeauthoryear {%
Rand%
}{%
Rand%
}{%
{\protect \APACyear {1971}}%
}]{%
rand1971objective}
\APACinsertmetastar {%
rand1971objective}%
\begin{APACrefauthors}%
Rand, W\BPBI M.%
\end{APACrefauthors}%
\unskip\
\newblock
\APACrefYearMonthDay{1971}{}{}.
\newblock
{\BBOQ}\APACrefatitle {Objective criteria for the evaluation of clustering
  methods} {Objective criteria for the evaluation of clustering
  methods}.{\BBCQ}
\newblock
\APACjournalVolNumPages{Journal of the American Statistical
  Association}{66}{336}{846--850}.
\PrintBackRefs{\CurrentBib}

\bibitem [\protect \citeauthoryear {%
Rasch%
}{%
Rasch%
}{%
{\protect \APACyear {1960/1980}}%
}]{%
rasch1960}
\APACinsertmetastar {%
rasch1960}%
\begin{APACrefauthors}%
Rasch, G.%
\end{APACrefauthors}%
\unskip\
\newblock
\APACrefYear{1960/1980}.
\newblock
\APACrefbtitle {Probabilistic models for some intelligence and attainment tests
  (ExpandEd Edition 1980)} {Probabilistic models for some intelligence and
  attainment tests (expanded edition 1980)}.
\newblock
\APACaddressPublisher{Chicago: Copenhagen}{Danish Institute for Education
  Research, The University of Chicago Press}.
\PrintBackRefs{\CurrentBib}

\bibitem [\protect \citeauthoryear {%
Robin%
\ \protect \BOthers {.}}{%
Robin%
\ \protect \BOthers {.}}{%
{\protect \APACyear {2011}}%
}]{%
Rpkg:pROC}
\APACinsertmetastar {%
Rpkg:pROC}%
\begin{APACrefauthors}%
Robin, X.%
, Turck, N.%
, Hainard, A.%
, Tiberti, N.%
, Lisacek, F.%
, Sanchez, J\BHBI C.%
\BCBL {}\ \BBA {} Müller, M.%
\end{APACrefauthors}%
\unskip\
\newblock
\APACrefYearMonthDay{2011}{}{}.
\newblock
{\BBOQ}\APACrefatitle {{pROC}: an open-source package for R and S+ to analyze
  and compare ROC curves} {{pROC}: an open-source package for r and s+ to
  analyze and compare roc curves}.{\BBCQ}
\newblock
\APACjournalVolNumPages{BMC Bioinformatics}{12}{1}{77}.
\newblock
\APACrefnote{{R} package version 0.3.0}
\PrintBackRefs{\CurrentBib}

\bibitem [\protect \citeauthoryear {%
J.~Rost%
}{%
J.~Rost%
}{%
{\protect \APACyear {1990}}%
}]{%
rost1990rasch}
\APACinsertmetastar {%
rost1990rasch}%
\begin{APACrefauthors}%
Rost, J.%
\end{APACrefauthors}%
\unskip\
\newblock
\APACrefYearMonthDay{1990}{}{}.
\newblock
{\BBOQ}\APACrefatitle {Rasch models in latent classes: An integration of two
  approaches to item analysis} {Rasch models in latent classes: An integration
  of two approaches to item analysis}.{\BBCQ}
\newblock
\APACjournalVolNumPages{Applied Psychological Measurement}{14}{3}{271--282}.
\PrintBackRefs{\CurrentBib}

\bibitem [\protect \citeauthoryear {%
J\BPBI E.~Rost%
\ \BBA {} Langeheine%
}{%
J\BPBI E.~Rost%
\ \BBA {} Langeheine%
}{%
{\protect \APACyear {1997}}%
}]{%
rost1997applications}
\APACinsertmetastar {%
rost1997applications}%
\begin{APACrefauthors}%
Rost, J\BPBI E.%
\BCBT {}\ \BBA {} Langeheine, R\BPBI E.%
\end{APACrefauthors}%
\unskip\
\newblock
\APACrefYear{1997}.
\newblock
\APACrefbtitle {Applications of latent trait and latent class models in the
  social sciences.} {Applications of latent trait and latent class models in
  the social sciences.}
\newblock
\APACaddressPublisher{}{Waxmann Publishing Co}.
\PrintBackRefs{\CurrentBib}

\bibitem [\protect \citeauthoryear {%
Schwarz%
}{%
Schwarz%
}{%
{\protect \APACyear {1978}}%
}]{%
schwarz1978estimating}
\APACinsertmetastar {%
schwarz1978estimating}%
\begin{APACrefauthors}%
Schwarz, G.%
\end{APACrefauthors}%
\unskip\
\newblock
\APACrefYearMonthDay{1978}{}{}.
\newblock
{\BBOQ}\APACrefatitle {Estimating the dimension of a model} {Estimating the
  dimension of a model}.{\BBCQ}
\newblock
\APACjournalVolNumPages{The annals of statistics}{6}{2}{461--464}.
\PrintBackRefs{\CurrentBib}

\bibitem [\protect \citeauthoryear {%
Sethuraman%
}{%
Sethuraman%
}{%
{\protect \APACyear {1991}}%
}]{%
sethuraman1994constructive}
\APACinsertmetastar {%
sethuraman1994constructive}%
\begin{APACrefauthors}%
Sethuraman, J.%
\end{APACrefauthors}%
\unskip\
\newblock
\APACrefYearMonthDay{1991}{}{}.
\newblock
{\BBOQ}\APACrefatitle {A Constructive Definition of {D}irichlet Priors} {A
  constructive definition of {D}irichlet priors}.{\BBCQ}
\newblock
\APACjournalVolNumPages{Statistics Sinica}{4}{2}{639-650}.
\PrintBackRefs{\CurrentBib}

\bibitem [\protect \citeauthoryear {%
Spiegelhalter%
, Best%
, Carlin%
\BCBL {}\ \BBA {} Van Der~Linde%
}{%
Spiegelhalter%
\ \protect \BOthers {.}}{%
{\protect \APACyear {2002}}%
}]{%
spiegelhalter2002bayesian}
\APACinsertmetastar {%
spiegelhalter2002bayesian}%
\begin{APACrefauthors}%
Spiegelhalter, D\BPBI J.%
, Best, N\BPBI G.%
, Carlin, B\BPBI P.%
\BCBL {}\ \BBA {} Van Der~Linde, A.%
\end{APACrefauthors}%
\unskip\
\newblock
\APACrefYearMonthDay{2002}{}{}.
\newblock
{\BBOQ}\APACrefatitle {Bayesian measures of model complexity and fit} {Bayesian
  measures of model complexity and fit}.{\BBCQ}
\newblock
\APACjournalVolNumPages{Journal of the Royal Statistical Society: Series B
  (Statistical Methodology)}{64}{4}{583--639}.
\PrintBackRefs{\CurrentBib}

\bibitem [\protect \citeauthoryear {%
Yang%
, O’Brien%
\BCBL {}\ \BBA {} Dunson%
}{%
Yang%
\ \protect \BOthers {.}}{%
{\protect \APACyear {2011}}%
}]{%
yang2011nonparametric}
\APACinsertmetastar {%
yang2011nonparametric}%
\begin{APACrefauthors}%
Yang, H.%
, O’Brien, S.%
\BCBL {}\ \BBA {} Dunson, D\BPBI B.%
\end{APACrefauthors}%
\unskip\
\newblock
\APACrefYearMonthDay{2011}{}{}.
\newblock
{\BBOQ}\APACrefatitle {Nonparametric Bayes stochastically ordered latent class
  models} {Nonparametric bayes stochastically ordered latent class
  models}.{\BBCQ}
\newblock
\APACjournalVolNumPages{Journal of the American Statistical
  Association}{106}{495}{807--817}.
\PrintBackRefs{\CurrentBib}

\end{thebibliography}


\begin{thebibliography}{}

\bibitem[Akaike, 1998]{akaike1998information}
Akaike, H. (1998).
\newblock Information theory and an extension of the maximum likelihood
  principle.
\newblock In {\em Selected papers of hirotugu akaike}, pages 199--213.
  Springer.

\bibitem[Bartolucci, 2007]{bartolucci2007class}
Bartolucci, F. (2007).
\newblock A class of multidimensional irt models for testing unidimensionality
  and clustering items.
\newblock {\em Psychometrika}, 72(2):141.

\bibitem[Bartolucci et~al., 2017]{bartolucci2017nonparametric}
Bartolucci, F., Farcomeni, A., and Scaccia, L. (2017).
\newblock A nonparametric multidimensional latent class irt model in a bayesian
  framework.
\newblock {\em Psychometrika}, 82(4):952--978.

\bibitem[de~Valpine et~al., 2017]{de2017programming}
de~Valpine, P., Turek, D., Paciorek, C.~J., Anderson-Bergman, C., Lang, D.~T.,
  and Bodik, R. (2017).
\newblock Programming with models: writing statistical algorithms for general
  model structures with {NIMBLE}.
\newblock {\em Journal of Computational and Graphical Statistics},
  26(2):403--413.

\bibitem[Ferguson, 1973]{ferguson1973bayesian}
Ferguson, T.~S. (1973).
\newblock A {B}ayesian analysis of some nonparametric problems.
\newblock {\em The Annals of Statistics}, 1(2):209--230.

\bibitem[Geng et~al., 2019a]{geng2019probabilistic}
Geng, J., Bhattacharya, A., and Pati, D. (2019a).
\newblock Probabilistic community detection with unknown number of communities.
\newblock {\em Journal of the American Statistical Association},
  114(526):893--905.

\bibitem[Geng et~al., 2019b]{geng2019bayesian}
Geng, J., Shi, W., and Hu, G. (2019b).
\newblock Bayesian nonparametric nonhomogeneous poisson process with
  applications to usgs earthquake data.
\newblock {\em arXiv preprint arXiv:1907.03186}.

\bibitem[Gnaldi et~al., 2016]{gnaldi2016multilevel}
Gnaldi, M., Bacci, S., and Bartolucci, F. (2016).
\newblock A multilevel finite mixture item response model to cluster examinees
  and schools.
\newblock {\em Advances in Data Analysis and Classification}, 10(1):53--70.

\bibitem[Hagenaars and McCutcheon, 2002]{hagenaars2002applied}
Hagenaars, J.~A. and McCutcheon, A.~L. (2002).
\newblock {\em Applied latent class analysis}.
\newblock Cambridge University Press.

\bibitem[Hu et~al., 2020]{hu2020bayesian}
Hu, G., Geng, J., Xue, Y., and Sang, H. (2020).
\newblock Bayesian spatial homogeneity pursuit of functional data: an
  application to the us income distribution.
\newblock {\em arXiv preprint arXiv:2002.06663}.

\bibitem[Ibrahim et~al., 2013]{ibrahim2013bayesian}
Ibrahim, J.~G., Chen, M.-H., and Sinha, D. (2013).
\newblock {\em Bayesian survival analysis}.
\newblock Springer Science \& Business Media.

\bibitem[Ishwaran and James, 2001]{ishwaran2001gibbs}
Ishwaran, H. and James, L.~F. (2001).
\newblock Gibbs sampling methods for stick-breaking priors.
\newblock {\em Journal of the American Statistical Association},
  96(453):161--173.

\bibitem[Lazarsfeld and Henry, 1968]{lazarsfeld1968latent}
Lazarsfeld, P.~F. and Henry, N.~W. (1968).
\newblock {\em Latent structure analysis}.
\newblock Houghton Mifflin Co.

\bibitem[Liu, 2019]{liu2019three}
Liu, X. (2019).
\newblock {\em Three Contributions to Latent Variable Modeling}.
\newblock PhD thesis, Columbia University.

\bibitem[Liu et~al., 2019]{liu2019comparison}
Liu, Y., Hu, G., Cao, L., Wang, X., and Chen, M.-H. (2019).
\newblock A comparison of monte carlo methods for computing marginal
  likelihoods of item response theory models.
\newblock {\em Journal of the Korean Statistical Society}.

\bibitem[Lunn et~al., 2000]{lunn2000winbugs}
Lunn, D.~J., Thomas, A., Best, N., and Spiegelhalter, D. (2000).
\newblock Winbugs-a bayesian modelling framework: concepts, structure, and
  extensibility.
\newblock {\em Statistics and computing}, 10(4):325--337.

\bibitem[Ma et~al., 2019]{ma2019bayesianspatial}
Ma, Z., Xue, Y., and Hu, G. (2019).
\newblock Heterogeneous regression models for clusters of spatial dependent
  data.
\newblock page arXiv:1907.02212.

\bibitem[Miller and Harrison, 2013]{miller2013simple}
Miller, J.~W. and Harrison, M.~T. (2013).
\newblock A simple example of {D}irichlet process mixture inconsistency for the
  number of components.
\newblock In {\em Advances in Neural Information Processing Systems}, pages
  199--206.

\bibitem[Miller and Harrison, 2018]{miller2018mixture}
Miller, J.~W. and Harrison, M.~T. (2018).
\newblock Mixture models with a prior on the number of components.
\newblock {\em Journal of the American Statistical Association},
  113(521):340--356.

\bibitem[Miyazaki and Hoshino, 2009]{miyazaki2009bayesian}
Miyazaki, K. and Hoshino, T. (2009).
\newblock A {B}ayesian semiparametric item response model with dirichlet
  process priors.
\newblock {\em Psychometrika}, 74(3):375--393.

\bibitem[Neal, 2000]{neal2000markov}
Neal, R.~M. (2000).
\newblock Markov chain sampling methods for {D}irichlet process mixture models.
\newblock {\em Journal of Computational and Graphical Statistics},
  9(2):249--265.

\bibitem[Pan and Huang, 2014]{pan2014bayesian}
Pan, J.-C. and Huang, G.-H. (2014).
\newblock Bayesian inferences of latent class models with an unknown number of
  classes.
\newblock {\em Psychometrika}, 79(4):621--646.

\bibitem[Pitman, 1995]{pitman1995exchangeable}
Pitman, J. (1995).
\newblock Exchangeable and partially exchangeable random partitions.
\newblock {\em Probability Theory and Related Fields}, 102(2):145--158.

\bibitem[Plummer et~al., 2003]{plummer2003jags}
Plummer, M. et~al. (2003).
\newblock Jags: A program for analysis of bayesian graphical models using gibbs
  sampling.
\newblock In {\em Proceedings of the 3rd international workshop on distributed
  statistical computing}, volume 124, page~10. Vienna, Austria.

\bibitem[{R Core Team}, 2013]{Rlanguage2013}
{R Core Team} (2013).
\newblock {\em R: A Language and Environment for Statistical Computing}.
\newblock R Foundation for Statistical Computing, Vienna, Austria.

\bibitem[Rand, 1971]{rand1971objective}
Rand, W.~M. (1971).
\newblock Objective criteria for the evaluation of clustering methods.
\newblock {\em Journal of the American Statistical Association},
  66(336):846--850.

\bibitem[Rasch, 1980]{rasch1960}
Rasch, G. (1960/1980).
\newblock {\em Probabilistic models for some intelligence and attainment tests
  (ExpandEd Edition 1980)}.
\newblock Danish Institute for Education Research, The University of Chicago
  Press, Chicago: Copenhagen.

\bibitem[Robin et~al., 2011]{Rpkg:pROC}
Robin, X., Turck, N., Hainard, A., Tiberti, N., Lisacek, F., Sanchez, J.-C.,
  and Müller, M. (2011).
\newblock {pROC}: an open-source package for r and s+ to analyze and compare
  roc curves.
\newblock {\em BMC Bioinformatics}, 12(1):77.
\newblock {R} package version 0.3.0.

\bibitem[Rost, 1990]{rost1990rasch}
Rost, J. (1990).
\newblock Rasch models in latent classes: An integration of two approaches to
  item analysis.
\newblock {\em Applied Psychological Measurement}, 14(3):271--282.

\bibitem[Rost and Langeheine, 1997]{rost1997applications}
Rost, J.~E. and Langeheine, R.~E. (1997).
\newblock {\em Applications of latent trait and latent class models in the
  social sciences.}
\newblock Waxmann Publishing Co.

\bibitem[Schwarz, 1978]{schwarz1978estimating}
Schwarz, G. (1978).
\newblock Estimating the dimension of a model.
\newblock {\em The Annals of Statistics}, 6(2):461--464.

\bibitem[Sethuraman, 1991]{sethuraman1994constructive}
Sethuraman, J. (1991).
\newblock A constructive definition of {D}irichlet priors.
\newblock {\em Statistics Sinica}, 4(2):639--650.

\bibitem[Spiegelhalter et~al., 2002]{spiegelhalter2002bayesian}
Spiegelhalter, D.~J., Best, N.~G., Carlin, B.~P., and Van Der~Linde, A. (2002).
\newblock Bayesian measures of model complexity and fit.
\newblock {\em Journal of the Royal Statistical Society: Series B (Statistical
  Methodology)}, 64(4):583--639.

\bibitem[Yang et~al., 2011]{yang2011nonparametric}
Yang, H., O’Brien, S., and Dunson, D.~B. (2011).
\newblock Nonparametric bayes stochastically ordered latent class models.
\newblock {\em Journal of the American Statistical Association},
  106(495):807--817.

\end{thebibliography}

\end{document}